\documentclass[journal=nalefd,manuscript=article]{achemso}

\usepackage{chemformula} % Formula subscripts using \ch{}
\usepackage[T1]{fontenc} % Use modern font encodings

\author{J. Bekaert}
\email{jonas.bekaert@uantwerpen.be}
\affiliation{%
 Department of Physics, University of Antwerp,
 Groenenborgerlaan 171, B-2020 Antwerp, Belgium
}%
\author{E. Khestanova}
\affiliation{%
National Graphene Institute, University of Manchester, Oxford Road, Manchester, United Kingdom, M13 9PL
}%
\alsoaffiliation{%
Department of Physics and Astronomy, University of Manchester, Oxford Road, Manchester, United Kingdom, M13 9PL
}%
\author{D. G. Hopkinson}
\affiliation{%
National Graphene Institute, University of Manchester, Oxford Road, Manchester, United Kingdom, M13 9PL
}%
\alsoaffiliation{%
Department of Materials, University of Manchester, Oxford Road, Manchester, United Kingdom, M13 9PL
}%
\author{J. Birkbeck}
\affiliation{%
National Graphene Institute, University of Manchester, Oxford Road, Manchester, United Kingdom, M13 9PL
}%
\alsoaffiliation{%
Department of Physics and Astronomy, University of Manchester, Oxford Road, Manchester, United Kingdom, M13 9PL
}%
\author{N. Clark}
\affiliation{%
National Graphene Institute, University of Manchester, Oxford Road, Manchester, United Kingdom, M13 9PL
}%
\alsoaffiliation{%
Department of Materials, University of Manchester, Oxford Road, Manchester, United Kingdom, M13 9PL
}%
\author{M. Zhu}
\affiliation{%
Department of Physics and Astronomy, University of Manchester, Oxford Road, Manchester, United Kingdom, M13 9PL
}%
\author{D. A. Bandurin}
\affiliation{%
Department of Physics and Astronomy, University of Manchester, Oxford Road, Manchester, United Kingdom, M13 9PL
}%
\author{R. Gorbachev}
\affiliation{%
National Graphene Institute, University of Manchester, Oxford Road, Manchester, United Kingdom, M13 9PL
}%
\alsoaffiliation{%
Department of Physics and Astronomy, University of Manchester, Oxford Road, Manchester, United Kingdom, M13 9PL
}%
\author{S. Fairclough}
\affiliation{%
Department of Materials, University of Manchester, Oxford Road, Manchester, United Kingdom, M13 9PL
}%
\author{Y. Zou}
\affiliation{%
Department of Materials, University of Manchester, Oxford Road, Manchester, United Kingdom, M13 9PL
}%
\author{M. Hamer}
\affiliation{%
National Graphene Institute, University of Manchester, Oxford Road, Manchester, United Kingdom, M13 9PL
}%
\author{D. J. Terry}
\affiliation{%
National Graphene Institute, University of Manchester, Oxford Road, Manchester, United Kingdom, M13 9PL
}%
\author{J. J. P. Peters}
\affiliation{%
School of Physics, University of Warwick, Coventry, CV4 7AL
}%
\author{A. M. Sanchez}
\affiliation{%
School of Physics, University of Warwick, Coventry, CV4 7AL
}%
 \author{B. Partoens}
\affiliation{%
 Department of Physics, University of Antwerp,
 Groenenborgerlaan 171, B-2020 Antwerp, Belgium
}
\author{S. J. Haigh}
\email{sarah.haigh@manchester.ac.uk}
\affiliation{%
National Graphene Institute, University of Manchester, Oxford Road, Manchester, United Kingdom, M13 9PL
}%
\alsoaffiliation{%
Department of Materials, University of Manchester, Oxford Road, Manchester, United Kingdom, M13 9PL
}%
 \author{M. V. Milo\v{s}evi\'{c}}
\email{milorad.milosevic@uantwerpen.be}
\affiliation{%
 Department of Physics, University of Antwerp,
 Groenenborgerlaan 171, B-2020 Antwerp, Belgium
}
\author{I. V. Grigorieva}
\email{irina.v.grigorieva@manchester.ac.uk}
\affiliation{%
National Graphene Institute, University of Manchester, Oxford Road, Manchester, United Kingdom, M13 9PL
}%
\alsoaffiliation{%
Department of Physics and Astronomy, University of Manchester, Oxford Road, Manchester, United Kingdom, M13 9PL
}%

\title{Enhanced superconductivity in few-layer TaS$_2$ due to healing by oxygenation}

\abbreviations{}

\begin{document}


\begin{tocentry}
\centering
\includegraphics[width=0.7\linewidth]{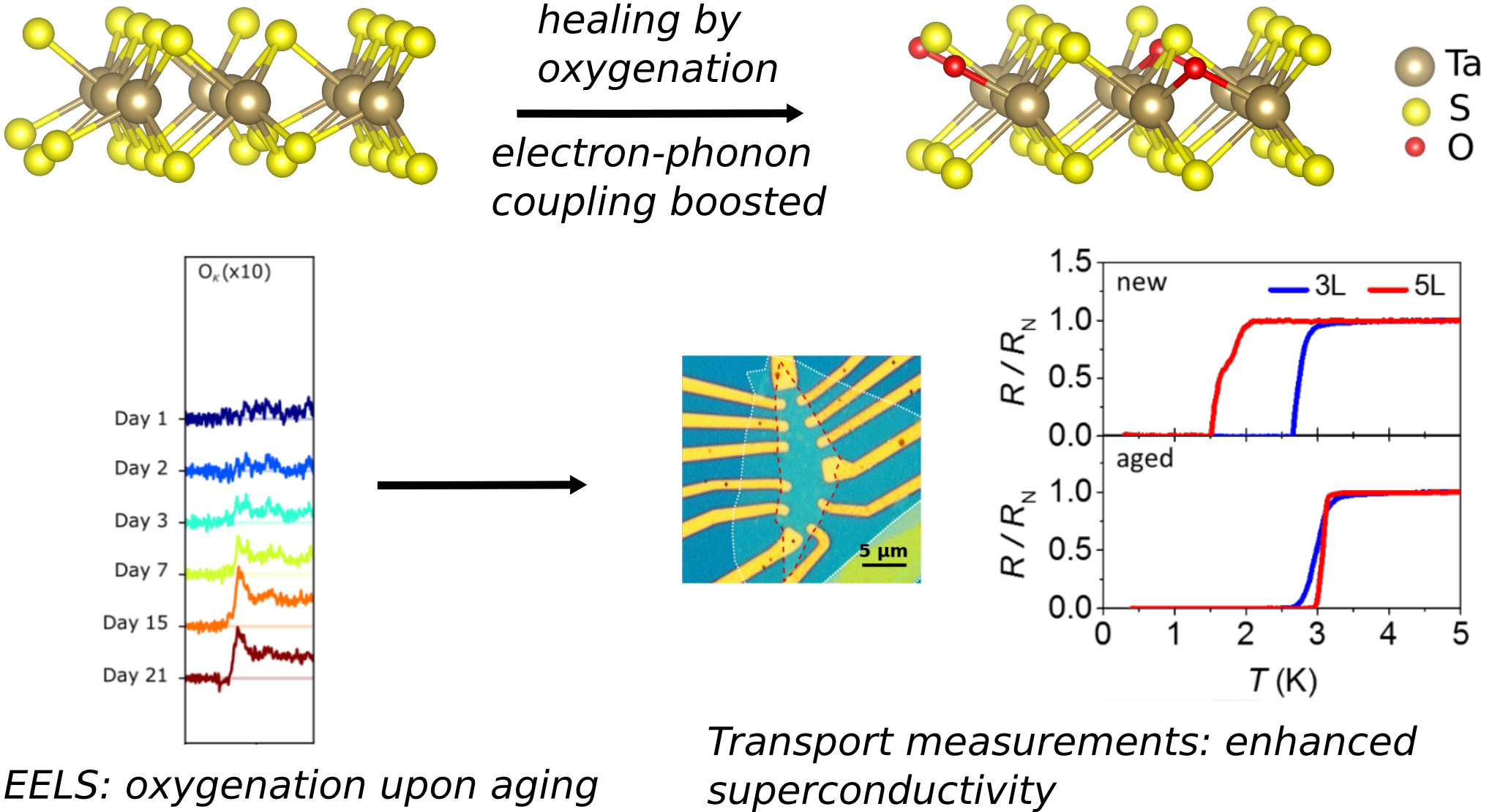}

\end{tocentry}

\begin{abstract}
When approaching the atomically thin limit, defects and disorder play an increasingly important role in the properties of two-dimensional materials. Superconductivity is generally thought to be vulnerable to these effects, but here we demonstrate the contrary in the case of oxygenation of ultrathin tantalum disulfide (TaS$_2$). Our first-principles calculations show that incorporation of oxygen into the TaS$_2$ crystal lattice is energetically favourable and effectively heals sulfur vacancies typically present in these crystals, thus restoring the carrier density to the intrinsic value of TaS$_2$. Strikingly, this leads to a strong enhancement of the electron-phonon coupling, by up to 80\% in the highly-oxygenated limit. Using transport measurements on fresh and aged (oxygenated) few-layer TaS$_2$, we found a marked increase of the superconducting critical temperature ($T_{\mathrm{c}}$) upon aging, in agreement with our theory, while concurrent electron microscopy and electron-energy loss spectroscopy confirmed the presence of sulfur vacancies in freshly prepared TaS$_2$ and incorporation of oxygen into the crystal lattice with time. Our work thus reveals the mechanism by which certain atomic-scale defects can be beneficial to superconductivity and opens a new route to engineer $T_{\mathrm{c}}$ in ultrathin materials.
\end{abstract}

\section{Introduction}

Shortly after the Bardeen-Cooper-Schrieffer (BCS) theory was established, Anderson applied it to a disordered system, proving that superconductivity in three-dimensional materials is robust against non-magnetic impurities (conserving both the energy gap and the critical temperature) \cite{ANDERSON195926}. However, for two-dimensional materials Anderson's analysis does not hold, and pair-breaking scattering due to disorder and impurities generally takes place\cite{0953-2048-30-1-013002}. A characteristic example are bismuth films undergoing a superconductor-to-insulator transition due to increased disorder in the thin limit \cite{PhysRevLett.62.2180}. Even with growth techniques for 2D materials having progressed enormously over recent years, defects can never be fully avoided. Hence, it is important to investigate how the effect of defects on superconductivity in 2D materials can be at least mitigated, or potentially even used to enhance the superconductivity. In doing so, one needs to go beyond the oversimplified picture that scattering from defects necessarily competes with the electron-phonon coupling responsible for Cooper-pair formation. As a more appropriate description, the effect of defects on the local electronic structure, the vibrational modes and the electron-phonon interaction need to be considered. In that respect, van der Waals materials are the ideal choice for a system to investigate, because they offer excellent control over sample thickness through growth or exfoliation with monolayer precision. Specifically, ultrathin transition metal dichalcogenides (TMDs) are currently attracting great interest owing to the rich interplay between structural, electronic and spin degrees of freedom they exhibit \cite{Manzeli2017}. In addition to superconductivity, ultrathin TMDs can host Ising spin textures due to strong spin-orbit coupling \cite{Lu1353,Saito2015,Xi2016,doi:10.1021/acs.nanolett.7b03026}, as well as charge density waves (CDWs) \cite{Xi2015,2053-1583-5-1-015006,Hovden11420,PhysRevLett.118.017002,2053-1583-4-4-041005,PhysRevB.98.035203}, the interaction of which provides a rich playground for new emergent phenomena.

In this work, we have focussed on few-layer crystals of the ground-state H-phase of tantalum disulfide (TaS$_2$) \cite{Manzeli2017}. The highly unexpected finding in 2016 of an increasing critical temperature ($T_{\mathrm{c}}$) with decreasing film thickness (from bulk value 0.6 K to 2.2 K for a five-layer film) sparked strong interest in ultrathin TaS$_2$ \cite{Navarro-Moratalla2016}. Subsequently, this trend was found to persist down to the monolayer, where $T_{\mathrm{c}}$ reaches 3 K\cite{delaBarrera2018}. Furthermore, random restacking of TaS$_2$ monolayers, eliminating inter-plane correlations, also produced a $T_{\mathrm{c}}$ of 3 K \cite{doi:10.1021/jacs.7b00216}. Similarly, an enhanced $T_{\mathrm{c}}$ of 3.4 K was found in bulk TaS$_2$ intercalated with organic molecules, which can be attributed to effectively monolayer behaviour, resulting from the intercalation, although charge transfer could also play a role \cite{doi:10.1063/1.430342}. The available explanations for the rise of $T_{\mathrm{c}}$ in ultrathin TaS$_2$ mainly rely on the weakening of the Coulomb interaction \cite{Navarro-Moratalla2016} and of the CDW state \cite{PhysRevB.98.035203,Wang2018}, though it remains puzzling that the behaviour of TaS$_2$ differs so strongly from that of other TMDs, such as NbSe$_2$, where $T_{\mathrm{c}}$ is observed to decrease in thinner samples \cite{Xi2016,Khestanova2018}.

In the investigations to date, the effect of atomic-scale defects in ultrathin TaS$_2$ has remained an almost entirely unexplored terrain. We therefore start by characterising the energetically favourable defects in few-layer TaS$_2$, and the associated changes in the electronic, vibrational and electron-phonon coupling properties and relate these to the superconducting state. The defects considered here are all associated with intrinsic lattice sites (either vacancies or substitutional defects), as opposed to the case of adatoms on monolayer superconductors whose additional electronic and vibrational states can also boost $T_{\mathrm{c}}$ \cite{PhysRevLett.123.077001}. First, we theoretically prove that oxygen substitution on sulfur lattice sites is energetically preferable to bare sulfur vacancies, and that oxygen effectively heals the lattice where sulfur vacancies were originally present, in the sense that the resulting carrier densities are indistinguishable from those in pristine TaS$_2$. Despite the equivalent carrier densities, oxygen substitution has a strong effect on electron-phonon coupling (EPC) as the latter increases by up to 80\% in the highly-oxygenated limit. The relevance of this effect to experimental observations is proven by scanning transmission electron microscopy (STEM) in combination with electron-energy loss spectroscopy (EELS) on exfoliated few-layer thick TaS$_2$ samples, which show a steady increase of the oxygen content in the lattice with sample aging. Concurrent transport measurements confirm the lattice healing, as the carrier densities return to their intrinsic values in the aged (oxygen-rich) samples and record a significant increase of $T_{\mathrm{c}}$ upon aging, which we link to the enhanced EPC due to oxygenation.

\section{Characterisation of defects in TaS$_2$}

The H-phase of TaS$_2$ comprises hexagonally symmetrical planes of Ta atoms sandwiched between two hexagonal sublattices of S atoms with trigonal prismatic coordination \cite{Manzeli2017}. Because of the weak van der Waals nature of the interactions between the different TaS$_2$ planes, we focus here on defects in a TaS$_2$ monolayer (ML). To find out which point defects in ultrathin TaS$_2$ are preferred energetically, we have calculated the formation energies of a selection of relevant defects, using density functional theory (DFT) as implemented in ABINIT \cite{Gonze20092582}, with inclusion of spin-orbit coupling (SOC). Specifically, we have considered either vacancy defects (sulfur or tantalum vacancies, $\mathrm{V}_{\mathrm{S}}$ and $\mathrm{V}_{\mathrm{Ta}}$ respectively) or substitutional defects (substitution of sulfur by tantalum, $\mathrm{Ta}_{\mathrm{S}}$, or by oxygen, $\mathrm{O}_{\mathrm{S}}$), within a $2 \times 2 \times 1$ supercell with fixed volume (computational details are provided in  Methods). 

In a metal, the formation energy of a point defect $\mathcal{D}$ is defined by
\begin{equation}
E_f\left(\mathcal{D}\right)=E_{\mathrm{tot}}\left(\mathcal{D}\right)-E_{\mathrm{tot}}\left(\mathrm{pristine}\right)+\sum_{\nu}n_{\nu}\mu_{\nu}~,
\label{eq:eformation}
\end{equation}
where the first two terms are the difference in total energy between the pristine and defect-containing supercells. The third term contains the chemical potentials of the exchanged atoms ($n_{\nu}=1$ or $n_{\nu}=-1$ for a removed or added atom, respectively), which can be written as $\mu_{\nu}=\mu_{\nu}^{\mathrm{elem}}+\Delta \mu_{\nu}$ with respect to the elemental phases of the exchanged atoms. In equilibrium conditions, variations in chemical potentials are limited by the heat of formation through $H_f\left(\mathrm{TaS}_2\right)=\Delta \mu_{\mathrm{Ta}}+2 \Delta \mu_{\mathrm{S}}$, providing a range of chemical potentials between sulfur-rich ($\Delta \mu_{\mathrm{S}}=0$, $\Delta \mu_{\mathrm{Ta}}=-2.93$ eV) and sulfur-poor ($\Delta \mu_{\mathrm{S}}=-1.47$ eV, $\Delta \mu_{\mathrm{Ta}}=0$), relating to different growth conditions. Since under typical experimental conditions oxygen is expected to be present in the form of O$_2$, we focus on oxygen-rich conditions ($\Delta \mu_{\mathrm{O}}=0$).

The obtained formation energies are summarized in Table \ref{tab:formation_energies1}, considering both S-rich and S-poor growth conditions. The results show that sulfur vacancies ($\mathrm{V}_{\mathrm{S}}$) are energetically favoured among the intrinsic defects but still require activation as $E_f\left(\mathrm{V}_{\mathrm{S}}\right)  >0$, even under S-poor conditions. This conclusion is independent of the supercell used ($2 \times 2$ vs. $3 \times 3$) or the film thickness (monolayer vs. bilayer). On the other hand, we find $E_f\left(\mathrm{O}_{\mathrm{S}}\right)<0$, i.e., oxygen substitutions on sulfur sites are energetically preferable regardless of the amount of available sulfur. This result indicates that, unlike e.g. NbSe$_2$ that has been reported to be resistant to oxygenation (at least in the absence of prior defects) \cite{Lin2019}, oxygenation of ultrathin TaS$_2$ samples can be expected to abundantly occur under realistic experimental conditions.

\begin{table}
\centering
\begin{tabular}{|c|c|c|c|c|}
\hline
Defect&No. of layers&Supercell&$E_f$ (eV), S-rich & $E_f$ (eV), S-poor\\ \hline \hline
$\mathrm{V}_{\mathrm{S}}$ & 1 & 2 $\times$ 2 & 2.60 & 1.14 \\ \hline
$\mathrm{V}_{\mathrm{S}}$ & 1 & 3 $\times$ 3 & 2.46 & 1.00 \\ \hline
$\mathrm{V}_{\mathrm{S}}$ & 2 & 2 $\times$ 2 & 2.58 & 1.12 \\ \hline
$\mathrm{Ta}_{\mathrm{S}}$ & 1 & 2 $\times$ 2 & 7.63 & 3.24 \\ \hline
$\mathrm{O}_{\mathrm{S}}$ & 1 & 2 $\times$ 2 & -1.21 & -2.67 \\ \hline
$\mathrm{V}_{\mathrm{Ta}}$ & 1 & 2 $\times$ 2 & 2.16 & 5.09 \\ \hline
\end{tabular}
\caption{Formation energies of point defects in atomically thin TaS$_2$.} 
\label{tab:formation_energies1}
\end{table}

\section{Carrier density and healing by oxygen}

\begin{figure}[t]
\centering
\includegraphics[width=1\linewidth]{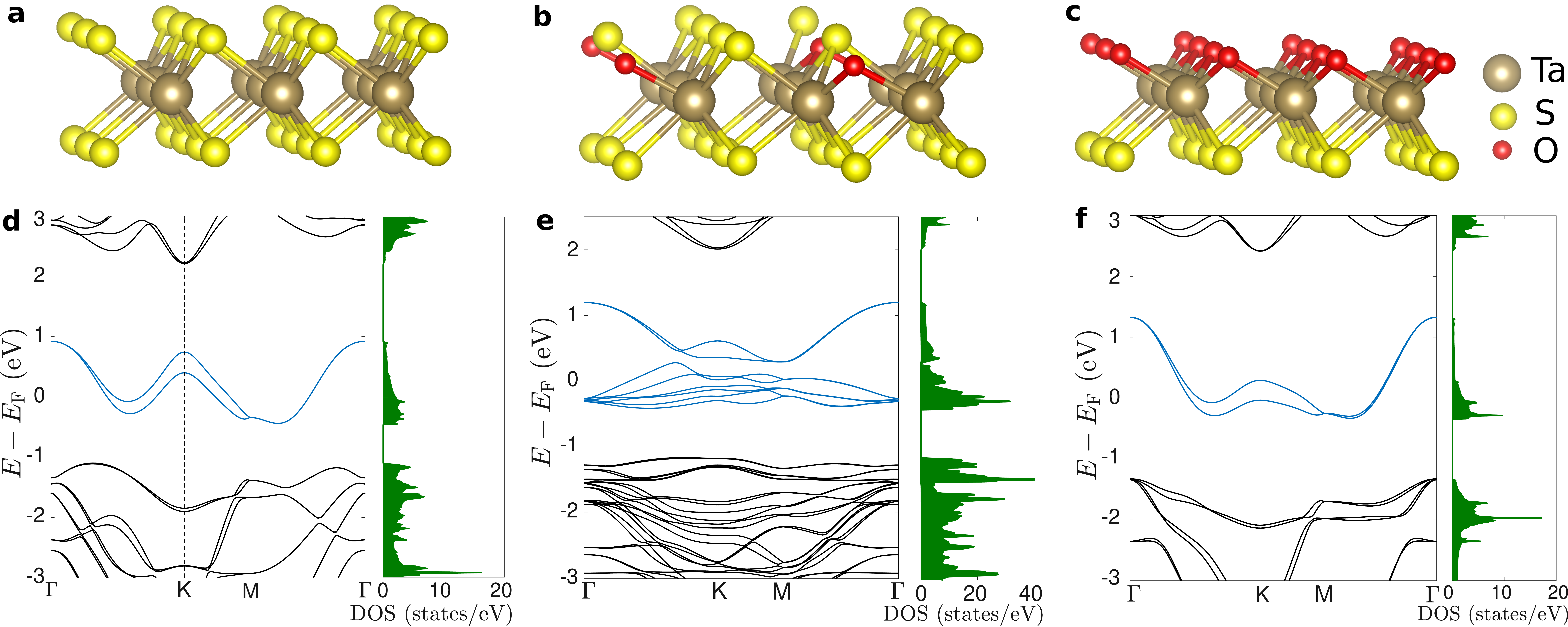}
\caption{Structural and electronic properties of ML TaS$_2$ upon oxygenation. Relaxed crystal structure of (\textbf{a}) pure ML TaS$_2$, (\textbf{b}) ML TaS$_2$ with one O substituting S per 4 unit cells (Ta$_4$S$_7$O) and (\textbf{c}) ML TaSO. Electronic band structure and density of states (DOS) of (\textbf{d}) TaS$_2$, (\textbf{e}) Ta$_4$S$_7$O and (\textbf{f}) TaSO.}
\label{fig:Fig3}
\end{figure}

To gain more insight in the effect of defects on the properties of ultrathin TaS$_2$, we investigated the electronic band structures and density of states (DOS) of pristine ML TaS$_2$ (cf.~Fig.~\ref{fig:Fig3}a), and ML TaS$_2$ containing point defects (obtained from DFT including SOC, within a $2 \times 2 \times 1$ supercell). Defects introduce significant changes in the DOS around the Fermi level ($E_{\mathrm{F}}$), which directly affects the electrical conductivity that can be experimentally probed. At very low temperatures (where the Fermi-Dirac distribution can be approximated by a step function), the carrier densities can be obtained by integrating the electronic DOS from the highest occupied electronic eigenvalue with zero DOS up to $E_{\mathrm{F}}$. In this way, we obtained the carrier densities listed in Table \ref{tab:carrier_densities}, from the calculated electronic structures (see Supporting Note 1 for further details). 

In the preceding analysis, we established that S vacancies are the most likely intrinsic defects to form. A sample containing S vacancies shows a higher carrier density ($1.51\cdot 10^{15}$ cm$^{-2}$) than pristine ML TaS$_2$ ($1.01\cdot 10^{15}$ cm$^{-2}$), in agreement with a recent experimental finding where an increased carrier density was found in TaS$_2$ monolayers containing vacancy clusters\cite{Peng2018}. However, when the S vacancies are filled by oxygen, leading to the structure depicted in Fig.~\ref{fig:Fig3}b, the carrier density is restored to its pristine value. This value, $1.01\cdot 10^{15}$ cm$^{-2}$, corresponds to one electron per Ta atom (i.e., per unit cell in the pristine case). The reason for this is that both in pristine TaS$_2$ and TaS$_2$ with an O$_{\mathrm{S}}$ substitution, the bands crossing $E_{\mathrm{F}}$ -- shown in Fig.~\ref{fig:Fig3}d,e -- are well-separated from the other bands and share one electron per Ta atom. 

Our results therefore directly support preferential lattice healing in TaS$_2$ by oxygen. The first healing effect is structural, as oxygen reduces the lattice symmetry breaking induced by the sulfur vacancy (as detailed in Supporting Note 1). On the electronic level, the carrier density is healed to its intrinsic values as oxygen fills a sulfur vacancy and this have profound consequences for superconductivity, as we demonstrate below.

\begin{table}[t]
\centering
\begin{tabular}{|c|c|}
\hline
System&Carrier density ($10^{15}$cm$^{-2}$)\\ \hline \hline
Pure&1.01 \\ \hline
$\mathrm{V}_{\mathrm{S}}$ & 1.51 \\ \hline
$\mathrm{Ta}_{\mathrm{S}}$ & 1.26 \\ \hline
$\mathrm{O}_{\mathrm{S}}$ & 1.01 \\ \hline
$\mathrm{V}_{\mathrm{Ta}}$ & 0.76 \\ \hline
\end{tabular}
\caption{Calculated carrier densities in ML TaS$_2$ without and with point defects.} 
\label{tab:carrier_densities}
\end{table}

\section{Enhanced superconductivity in oxygenated TaS$_2$}

To quantify the effect of oxygen on the superconducting properties of TaS$_2$, we have calculated the phonon modes and electron-phonon coupling (EPC) in both pristine and oxygenated ML TaS$_2$ using density functional perturbation theory (DFPT) \cite{PhysRevLett.58.1861,PhysRevLett.68.3603,PhysRevLett.69.2819} [see Methods for details].  
In the oxygenated case we consider both an intermediate and a high level of oxygenation, with unit formulas of Ta$_4$S$_7$O and TaSO, respectively, corresponding to the structures depicted in Fig.~\ref{fig:Fig3}b,c (in the latter case an entire plane of S is replaced by O). In both cases, substitution of sulfur with oxygen has a profound effect on the electronic band structure compared to TaS$_2$ with S vacancies (or other vacancy defects), see Supporting Fig.~S1. In the highly-oxygenated limit, TaSO (Fig.~\ref{fig:Fig3}f) the band structure is very similar to that of pristine TaS$_2$, consistent with the healing effect due to oxygen. This is to be expected, as the calculated TaSO structure preserves the vertical alignment of S and O characteristic of the H-phase  (see Supporting Note 1 for further discussion). In the partially oxygenated case, the `healing' of the band structure is not complete, with a few more bands present at $E_{\mathrm{F}}$, but it is nevertheless much closer to the pristine TaS$_2$ compared to unsaturated sulfur vacancies or other substitutions (cf. Supporting Fig.~S2).     

\begin{figure}
\centering
\includegraphics[width=0.85\linewidth]{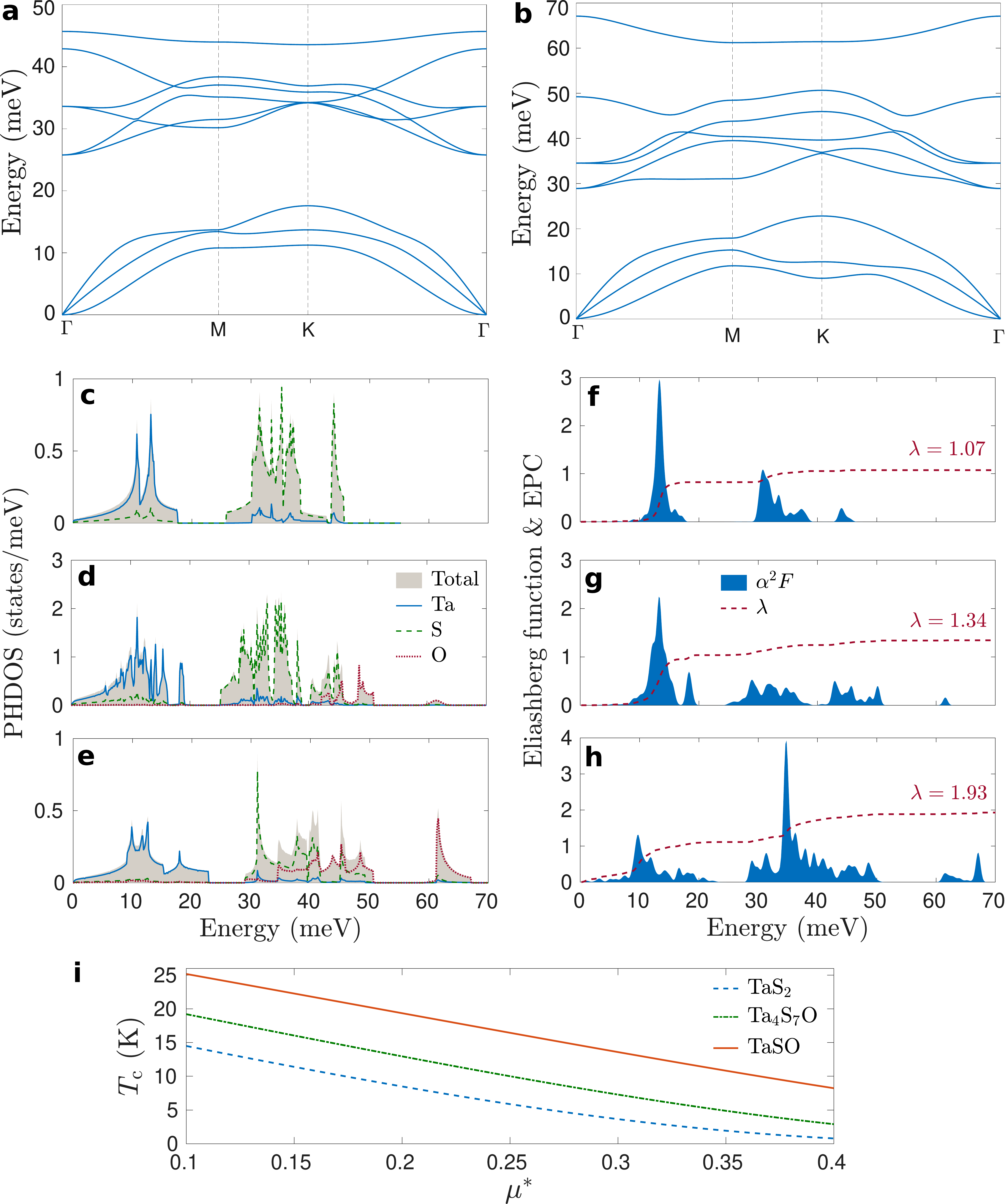}\\
\caption{Phonons and electron-phonon interaction in pure and oxygenated TaS$_2$. Phonon band structure of (\textbf{a}) ML TaS$_2$ and (\textbf{b}) ML TaSO. Phonon DOS (PHDOS) of (\textbf{c})  ML TaS$_2$, (\textbf{d}) ML Ta$_4$S$_7$O and (\textbf{e}) ML TaSO. Eliashberg function $\alpha^2F$ plus electron-phonon coupling constant $\lambda$ of (\textbf{f})  ML TaS$_2$, (\textbf{g}) ML Ta$_4$S$_7$O and (\textbf{h}) ML TaSO. (\textbf{i}) Critical temperature of ML TaS$_2$, Ta$_4$S$_7$O and TaSO as a function of Coulomb pseudopotential $\mu^*$.}
\label{fig:Fig4}
\end{figure}

The calculated phonon band structures and phonon DOS's (PHDOS) of these three cases are shown in Figs.~\ref{fig:Fig4}c--e. The atom-resolved phonon DOS shows that, because of the large difference in atomic mass, the phonon modes due to Ta (acoustic modes at lower energies) and those due to the chalcogen atoms (optical modes at higher energies) are quite well-resolved. As expected, the main changes due to $\mathrm{O}_{\mathrm{S}}$ substitution thus occur in the chalcogen-related optical modes. 

To evaluate the influence of these changes in the phonon modes on the superconducting properties we have used Eliashberg theory, which provides a quantitative description of phonon-mediated superconductivity beyond the BCS theory \cite{Eliashberg1960,Eliashberg1961,RevModPhys.89.015003}. Here, the central role is played by the Eliashberg function, 
\begin{equation}
\alpha^2F(\omega)=\frac{1}{N_{\mathrm{F}}}\sum_{\nu,\textbf{k},\textbf{q}}\left|g_{\textbf{k},\textbf{k}+\textbf{q}}^{\nu}\right|^2 \delta\left( \omega-\omega_{\textbf{q}}^{\nu}\right)\delta\left(\epsilon_{\textbf{k}} \right)\delta\left(\epsilon_{\textbf{k}+\textbf{q}} \right)~,
\end{equation}
which depends on the electronic DOS at $E_{\mathrm{F}}$ ($ N_{\mathrm{F}}$), the electron-phonon scattering matrix elements ($g_{\textbf{k},\textbf{k}+\textbf{q}}^{\nu}$), the phonon dispersion ($\omega_{\textbf{q}}^{\nu}$), and the electronic dispersion ($\epsilon_{\textbf{k}}$), all obtained from DFT and DFPT. From the Eliashberg function the EPC constant can be obtained as $
\lambda=2\int_0^\infty \alpha^2F(\omega)\omega^{-1}d\omega$. 

The results for pure and oxygenated TaS$_2$ are shown in Fig.~\ref{fig:Fig4}f--h. Prior to oxygenation (Fig.~\ref{fig:Fig4}f), the predominant contribution to the coupling arises from the Ta-based acoustic modes (associated with the formation of the CDW instability \cite{PhysRevB.95.235121}). Upon substitution by oxygen of one sulfur atom within a $2 \times 2 \times 1$ supercell (i.e., unit formula Ta$_4$S$_7$O, shown in Fig.~\ref{fig:Fig4}g), coupling to the high-energy oxygen-based phonon modes emerges, with a total enhancement of the EPC by 25\% with respect to the pristine case. Finally, in the highly-oxygenated limit (Fig.~\ref{fig:Fig4}h), the maximal EPC has radically shifted to the chalcogen-based optical modes (of mixed S and O character). This enhanced coupling to the optical modes likely results from the shorter Ta-O bond length (as shown in Fig.~\ref{fig:Fig3}c), strengthening the interaction between the electronic states with dominant Ta-d character and O-based phonon modes. The overall result is an enhancement of the EPC constant by 80\% in the highly-oxygenated limit, despite the nearly constant electronic DOS before and after oxygenation. 

We evaluated the resulting change in the superconducting $T_{\mathrm{c}}$ using the McMillan-Allen-Dynes formula \cite{PhysRev.167.331,PhysRevB.12.905}, as summarised in Table \ref{tab:EPH}. The obtained $T_{\mathrm{c}}$ of ML TaS$_2$ (11 K) is clearly overestimated with respect to the experimental value of 3 K \cite{PhysRevB.98.035203,delaBarrera2018}. This lower value of $T_{\mathrm{c}}$ could result from the competition of superconductivity with the CDW phase, but can also be related, at least partially, to an enhanced Coulomb repulsion between the Cooper-pair electrons. To this end, we investigated the dependence of $T_{\mathrm{c}}$ on the Coulomb pseudopotential $\mu^*$ \cite{PhysRev.125.1263}, as shown in Fig.~\ref{fig:Fig4}i. This analysis shows that to obtain the experimental value of $T_{\mathrm{c}}$ for ML TaS$_2$ requires $\mu^*=0.32$, well beyond the common range of 0.1--0.15 \cite{ALLEN19831}. However, crucially for this work, the enhancement of $T_{\mathrm{c}}$ due to oxygenation (both at lower and higher concentrations) is robust over the whole $\mu^*$ range. 

\begin{table}
\centering
\begin{tabular}{|c|c|c|}
\hline
Compound&$\lambda$&$T_{\mathrm{c}}(\mu^*=0.15)$ (K)\\ \hline \hline
TaS$_2$& 1.07 &  11 \\ \hline
Ta$_4$S$_7$O& 1.34 &  16 \\ \hline
TaSO& 1.93 &  22 \\ \hline
\end{tabular}
\caption{The calculated EPC constant $\lambda$ and the resulting critical temperature of pure and oxygenated ML TaS$_2$.} 
\label{tab:EPH}
\end{table}

\section{ADF-STEM and EELS analysis}

To relate our theory findings to the presence of defects in ultrathin samples of TaS$_2$, we prepared mono-, bi- and few-layer TaS$_2$ crystals and studied them using atomic resolution annular dark field scanning transmission electron microscopy (ADF-STEM) and electron-energy loss spectroscopy (EELS). Atomically thin TaS$_2$ is known to be unstable in ambient conditions. Therefore, to minimise degradation, the crystals were mechanically exfoliated from bulk TaS$_2$ in the inert atmosphere of a glovebox and encapsulated with monolayer graphene following the procedure reported in ref.~\citenum{Clark2018}. The encapsulation is known to provide reliable protection from moisture and oxygen in ambient conditions as long as the resulting graphene-TMD-graphene (or hBN-TMD-hBN) stack exhibits so-called self-cleaning \cite{doi:10.1021/acs.nanolett.5b00648,Kretinin2014}, i.e., strong adhesion between atomically thin crystals squeezes out adsorbed water and hydrocarbons into sub-micrometer-size pockets (bubbles), leaving the remaining interfaces atomically sharp and largely free of contaminants. In the case of TaS$_2$, however, we found self-cleaning to be poor (cf. also Supporting Fig.~S6) so that a thin contamination layer often remained despite encapsulation, effectively providing a reservoir of carbon, hydrogen and oxygen that could diffuse along the graphene-TaS$_2$ interface and react with the encapsulated TaS$_2$.

\begin{figure}
\centering
\includegraphics[width=\linewidth]{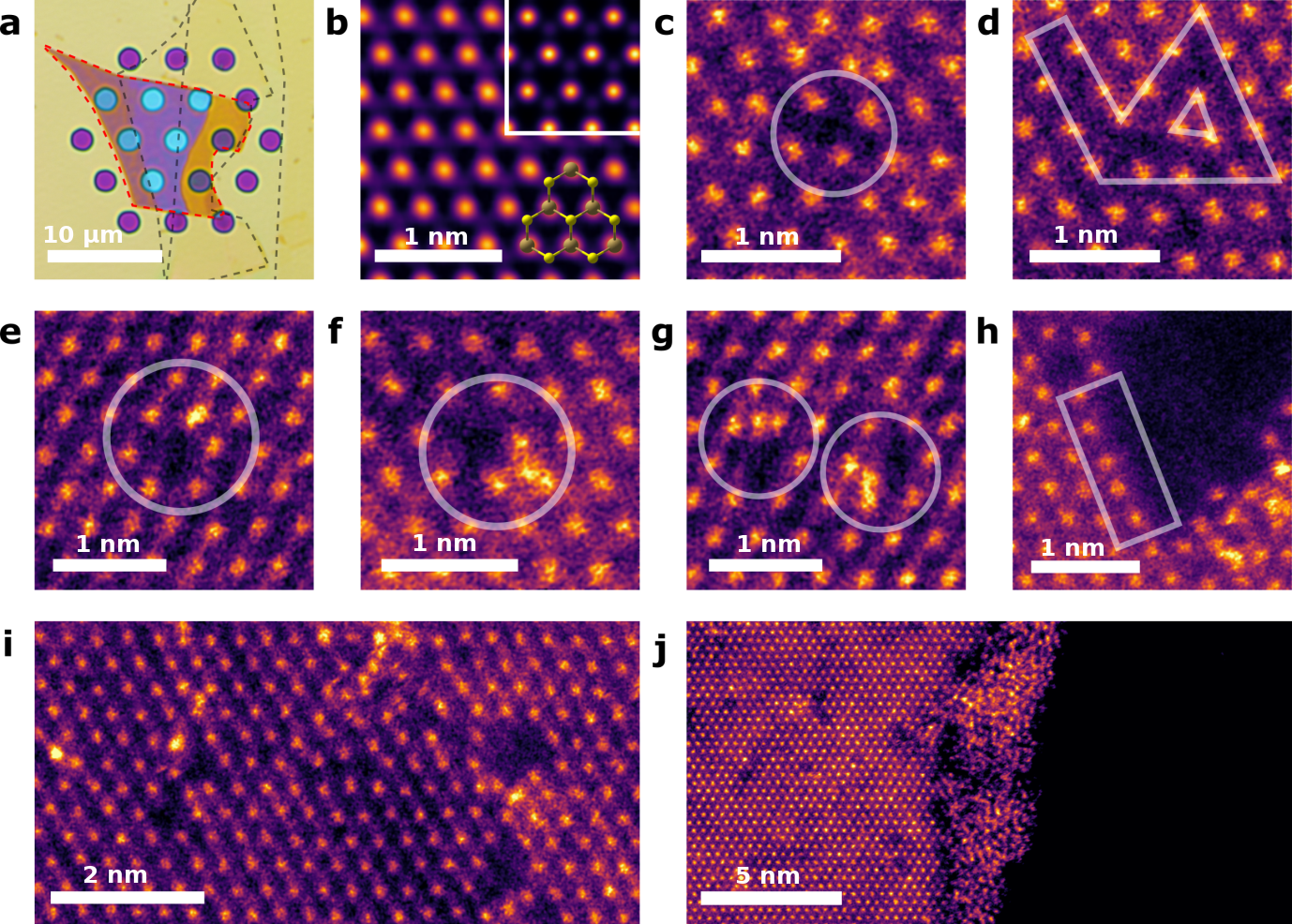}
\caption{ADF-STEM imaging of defects in atomically thin TaS$_2$. (\textbf{a}) Typical sample of  graphene-encapsulated TaS$_2$, with the TaS$_2$ flake outlined in red and the upper and lower graphene encapsulation outlined in grey. (\textbf{b}) Drift-corrected, time-averaged image of graphene-encapsulated monolayer TaS$_2$, with the top-right inset showing an image simulation and the bottom-right an overlaid atomic model (Ta atoms are gold, S atoms are yellow).The time averaging across 100 high speed images improves the signal-to-noise ratio but also averages the contrast variations due to mobile defects. (\textbf{c,d}) S vacancy point and line defects in monolayer TaS$_2$, respectively. (\textbf{e-g}) Ta vacancy and adatom defects in monolayer TaS$_2$, with Ta adatoms located at the adjacent Ta (e), hollow (f), bridging (g, left), and S (g, right) sites. (\textbf{h}) Atomically-flat edge of a nm-scale pore in a monolayer TaS$_2$. (\textbf{i}) Defect clusters in monolayer TaS$_2$. (\textbf{j}) Amorphous edge in bilayer TaS$_2$.}
\label{fig:STEM}
\end{figure}

Fig.~\ref{fig:STEM} shows the results of high-resolution ADF-STEM imaging of freshly prepared mono- and bilayer TaS$_2$ encapsulated with graphene. Tantalum atoms appear very bright due to the sensitivity of the technique to the atomic number, with sulphur positions visible as low-intensity spots between Ta atoms. Ta vacancies ($\mathrm{V}_{\mathrm{Ta}}$) are seen clearly in all images, while the much lower scattering cross-section of sulfur makes identification of $\mathrm{V}_{\mathrm{S}}$ non-trivial even for monolayer crystals (see Supporting Fig. S3 and the related discussion). Nevertheless, the absence of S atoms can be identified by a reduced ADF-STEM intensity and the displacement of the adjacent Ta atoms -- these are clearly visible in Fig.~\ref{fig:STEM}d,i. Importantly, large numbers of S vacancies, as well as some Ta vacancies were identified already in the first high resolution scans, i.e., they are not the result of electron beam damage, which supports our expectation from theory that low $\mathrm{V}_{\mathrm{S}}$ formation energies should result in an abundance of such defects even in carefully prepared, mechanically exfoliated and encapsulated samples. 

Under electron beam irradiation, new defects at both S and Ta sites were seen to form, migrate to nearby lattice sites, and heal back to a contiguous lattice. Sulfur vacancies in particular were prone to migrate, heal or coalesce into line defects (Fig.~\ref{fig:STEM}d), a structure previously seen for vacancy defects in several other TMDs, such as MoS$_2$ \cite{PhysRevB.88.035301,Wang2016}.
In addition to vacancies and line defects, the edges of the encapsulated crystals (where graphene-TaS$_2$-graphene stack transition to graphene-graphene) were amorphous, typically reaching 1--3 nm into the crystal. We speculate that this edge amorphization is the result of chemical reactions with hydrocarbon residues and water that are present despite encapsulation due to poor self-cleaning. 

\begin{figure}
\centering
\includegraphics[width=\linewidth]{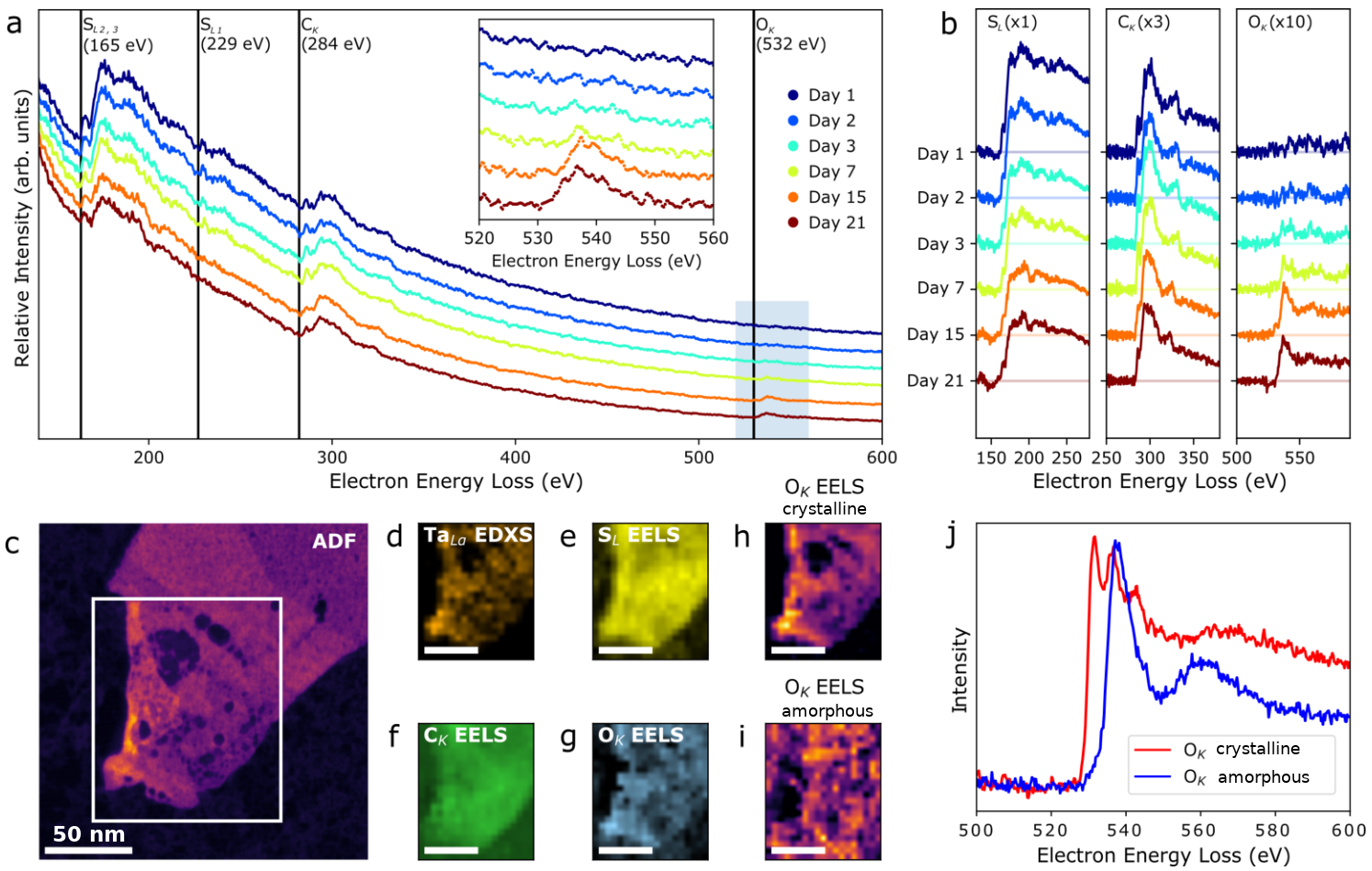}
\caption{Time-dependent compositional variation in TaS$_2$ using EELS. 0 -- 3 weeks: (\textbf{a}) High-loss spectra from an encapsulated 3-ML region, with detail of the oxygen-K edge as an inset. (\textbf{b}) Detail of extracted sulfur-L, carbon-K, and oxygen-K edges. 7 months: (\textbf{c}) ADF-STEM of TaS$_2$ 7 months after fabrication. (\textbf{d})--(\textbf{g}) Intensity maps for the extracted signals of (\textbf{d}) tantalum-L$\alpha$ EDXS peak, (\textbf{e}) sulfur-L EELS edges, (\textbf{f}) carbon-K EELS edge, and (\textbf{g}) oxygen-K EELS edge. (\textbf{h}) and (\textbf{i}) Maps of two oxygen-K edge components, associated with (\textbf{h}) crystalline and (\textbf{i}) amorphous environments from the region indicated in (\textbf{c}). (\textbf{j}) Comparison of the isolated oxygen-K edge components corresponding to the crystalline and amorphous oxides.}
\label{fig:EELS}
\end{figure}

To investigate the evolution of the chemical composition of the encapsulated TaS$_2$ crystals with time, we performed STEM-EELS analysis over the course of 3 weeks and on a crystal aged for 7 months. To limit the effect of beam-induced contamination, consecutive measurements were made on similar but not identical regions of the sample. As demonstrated in Fig.~\ref{fig:EELS}a,b, the oxygen content in the TaS$_2$ sample clearly increases with time, from less than 1 at\% oxygen in the first 3 weeks (S:O ratio of $\sim$100:1) to  $\sim$50\% of sulfur replaced by oxygen after 7 months of storage (see Supporting Note 3 for details of the analysis). To exclude the possibility that the detected oxygen is present in the hydrocarbon contamination layer rather than being associated with TaS$_2$, additional EELS measurements were carried out in a neighbouring region of the same sample containing just the two graphene encapsulating layers. This showed a markedly smaller increase in oxygen content, demonstrating that the oxygen increase in Fig.~\ref{fig:EELS}a occurred predominantly within the TaS$_2$ crystal lattice. 

To elucidate the form in which oxygen is present after 7 months of storage, we used independent component analysis of the EEL spectrum image/map. This revealed the presence of two separate oxygen-containing components: component 1, responsible for $\sim$70\% of the strength of the oxygen signal and  collocated with the Ta and S atoms, has a spectral shape  often associated with crystalline transition metal oxides (showing a strong pre-peak, see  Fig.~\ref{fig:EELS}j), and is similar to the spectra for Ta$_2$O$_5$ reported in Refs.~\citenum{C2CP41786C,Song2017}. We therefore assign this peak to oxygen incorporated into the TaS$_2$ lattice, presumably as substitution for sulfur atoms as predicted by theory. The peak associated with component 2, found at a higher energy, Fig.~\ref{fig:EELS}j, is a typical O signature from contamination found on the surfaces of 2D crystals, e.g. where oxygen or water are adsorbed in hydrocarbons inevitably present on or around our TaS$_2$ crystals, as explained above. Furthermore, our STEM-EELS analysis showed that the rate of oxygenation depends on the sample thickness. For example, we found  a much higher relative increase in oxygen content for the trilayer region in Fig.~\ref{fig:STEM}a compared to similarly encapsulated $\sim$15 nm-thick crystals -- see Fig. S5 in the Supporting Information. 

The above findings provide strong support for our theory predictions: the presence of large numbers of vacancies, particularly S vacancies in freshly prepared samples, and at least partial `healing' of the defects with oxygen, as the latter is seen to be incorporated into TaS$_2$ lattice. 

\section{Transport measurements}

\begin{figure}
\centering
\includegraphics[width=1\linewidth]{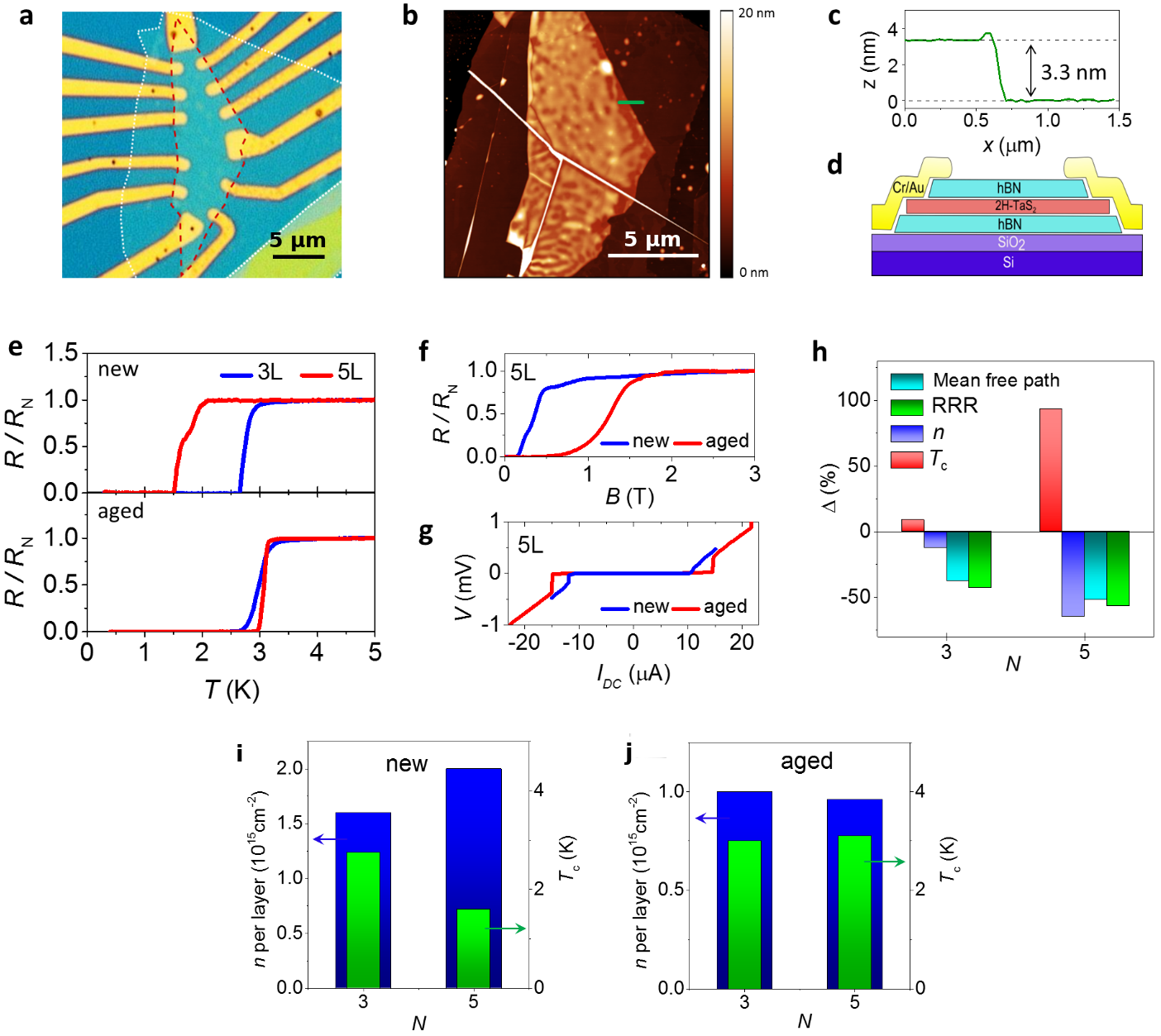}
\caption{Enhancement of the superconducting properties in aged few-layer TaS$_2$ based on transport measurements. (\textbf{a}) Optical image of a 5-ML device, where the red dashed line indicates the TaS$_2$ flake and the white line the encapsulating hBN, (\textbf{b}) AFM image of the encapsulated flake, (\textbf{c}) step profile along the green line in (\textbf{b}), and (\textbf{d}) schematic side view of a transport device.  (\textbf{e}) Normalised resistance as a function of temperature for new and aged (after 3 weeks) devices, (\textbf{f}) perpendicular magnetic-field dependence of superconductivity in the 5-ML device at $T = 0.3$ K, (\textbf{g}) zero-field critical current at $T = 0.3$ K for the same device. (\textbf{h}) Relative changes over time in the carrier concentration $n$, mean free path $l$, RRR and $T_{\mathrm{c}}$. (\textbf{i},\textbf{j}) Carrier concentration per layer and $T_{\mathrm{c}}$ for new and aged 3-ML and 5-ML devices.}
\label{fig:Transport}
\end{figure}

To verify the theoretically predicted enhancement of the superconducting properties due to oxygenation, we carried out transport measurements on TaS$_2$ samples with different numbers of layers  $N$ ($N=3$ and $N=5$), both freshly prepared (within a few days after fabrication) and stored in a desiccator for several weeks (we refer to the latter as `aged'). The devices were made using an exfoliation and encapsulation procedure very similar to that used for STEM-EELS samples, except that few-layer TaS$_2$ flakes were encapsulated by hexagonal boron nitride (hBN) rather than graphene. The details can be found in ref.~\citenum{Khestanova2018} and Supporting Note 4. The thickness of each sample was identified using optical contrast and the smallest step height measured over the TaS$_2$ edge by AFM (cf. Fig.~\ref{fig:Transport}c and Supporting Fig.~S6). Electrical contacts to TaS$_2$ were defined  after encapsulation using e-beam lithography. To this end, top hBN layer was selectively etched using CHF$_3$ plasma, uncovering the surface of TaS$_2$ that was then immediately metallized by evaporating Cr/Au layer to form the electrodes (see Fig.~\ref{fig:Transport}d). Further etching (e.g. to shape Hall bars) was avoided to maintain the integrity of encapsulation and prevent access of oxidants via the exposed edge of TaS$_2$. 

The superconducting characteristics ($T_{\mathrm{c}}$, critical magnetic field $H_{\mathrm{c2}}$) of freshly prepared samples were found to be very similar to those reported in literature for TaS$_2$ crystals of the same thickness\cite{Navarro-Moratalla2016,PhysRevB.98.035203,delaBarrera2018}, see Fig.~\ref{fig:Transport}e. As found previously, $T_{\mathrm{c}}$ is higher for thinner crystals, in our case $T_{\mathrm{c}}\approx1.7$ K for a 5-layer device and $T_{\mathrm{c}}\approx2.7$ K for the trilayer (here $T_{\mathrm{c}}$ is determined as the mid-point of the superconducting transition on $R(T)$). Also in agreement with previous reports, the carrier densities $n$ extracted from our Hall measurements in van der Pauw geometry \cite{vdPauw1958} (Supporting Note 4 and Fig.~\ref{fig:Transport}i) significantly exceeded the known values for pristine TaS$_2$, indicating the presence of vacancy defects \cite{Peng2018}. We note that the experimental values of $n\sim 1.5\cdot 10^{15}$ cm$^{-2}$/layer for both our devices are in excellent agreement with our calculations for TaS$_2$ monolayers with high-density of S vacancies [cf.~Table \ref{tab:carrier_densities}].

After several weeks of storage the measurements were repeated using the same pairs of contacts. As predicted, we found that $T_{\mathrm{c}}$ markedly increased, saturating at $T_{\mathrm{c}}\approx3$ K for both thicknesses of TaS$_2$ (see Fig.~\ref{fig:Transport}e). The perpendicular $H_{c2}$ and the critical current also increased, as shown in Fig.~\ref{fig:Transport}f,g, indicating an overall enhancement of the superconducting properties due to sample aging. The superconducting transition for both devices also became noticeably sharper, indicating that aging made the devices more uniform.
In contrast to the enhancement of  $T_{\mathrm{c}}$ and better uniformity, the normal state transport properties showed marked degradation, with the residual resistance ratio $\mathrm{RRR}=R(300~ \mathrm{K})/R(10~ \mathrm{K})$ and the mean-free path $l$ decreasing by as much as 50\% (Fig.~\ref{fig:Transport}h and Supporting Fig.~S7).  At the same time, the carrier density in both aged devices decreased to the intrinsic value for TaS$_2$ ($1 \cdot 10^{15}$ cm$^{-2}$/layer, see Fig.~\ref{fig:Transport}j), independent of sample thickness. Taken together, comparison of the relative change ($\Delta$) in the normal-state transport characteristics and $T_{\mathrm{c}}$ (Fig.~\ref{fig:Transport}h) shows an anti-correlation between the two, with the largest reduction in $n$, RRR and $l$ corresponding to the largest increase in $T_{\mathrm{c}}$ (further details are provided in Supporting Note 4). 

Although counter-intuitive,  the above observations are in excellent agreement with our theoretical predictions. The `intrinsic' per-layer carrier densities and sharp superconducting transitions in the aged samples, in combination with the steady increase in the oxygen content revealed by EELS and its incorporation into the TaS$_2$ lattice, provide experimental evidence of the predicted healing of the lattice by oxygenation. In turn, this can explain the enhancement of $T_{\mathrm{c}}$ and the critical field after aging, due to the enhanced EPC inherent to oxygenation, as found in our calculations.

\section{Conclusions}

The above results demonstrate that, contrary to widely held assumptions, superconductivity in two-dimensional materials can benefit from the effect of oxygenation and establish few-layer TaS$_2$ as a system where 2D superconductivity is enhanced by oxygenation.

This effect is distinctly different from that described for monolayer TaS$_2$ samples containing chemically etched nanopores \cite{Peng2018}, where elevated critical temperatures (maximum onset of 3.6 K) were attributed to an enhancement of the carrier density. From this, Peng \textit{et al.} concluded that the density of states at the Fermi level must be equally enhanced, implying an increased electron-phonon interaction. However, as demonstrated by our transport measurements (Fig.~\ref{fig:Transport}h--j), there is no straightforward relation between the carrier density and $T_{\mathrm{c}}$ for few-layer TaS$_2$. While for freshly prepared samples we also found larger per-layer $n$ compared to the intrinsic (theoretical) value and attribute it to the presence of defects, aging and oxygenation lead to a decrease in $n$ but nevertheless an increase in $T_{\mathrm{c}}$. Furthermore, our calculations show that vacancies, particularly sulfur vacancies, do not lead to a significant enhancement of the DOS at $E_{\mathrm{F}}$ (see Supporting Fig.~S2) and therefore cannot explain the large increase in $T_{\mathrm{c}}$ observed in the presence of defects.   

Another mechanism of $T_{\mathrm{c}}$ enhancement in ultrathin TaS$_2$ crystals compared to bulk TaS$_2$ was suggested in refs.~\citenum{Navarro-Moratalla2016,PhysRevB.98.035203}, where it was interpreted as being due to weakening of the CDW state for crystals with $N\leq 5$, even though there is still no direct experimental evidence of CDW weakening in few-layer TaS$_2$ \cite{PhysRevB.98.035203} and the microscopic mechanism of its effect on superconductivity remains under discussion. The critical temperatures for our freshly prepared 3- and 5-layer devices are in excellent agreement with $T_{\mathrm{c}}$ values for the same sample thicknesses in ref.~\citenum{PhysRevB.98.035203}. This suggests a common origin of $T_{\mathrm{c}}$ enhancement in both studies, possibly related to defects, possibly to a modification of the CDW state and its competition with superconductivity (we note that atomically thin crystals in ref.~\citenum{PhysRevB.98.035203} and in this work were prepared using the same mechanical exfoliation and encapsulation method). Importantly, however, the nearly 50\% further increase of $T_{\mathrm{c}}$ as a result of oxygenation (for $N=5$) and a smaller but still appreciable increase for $N=3$ is a new effect, distinct from the previously discussed mechanisms of $T_{\mathrm{c}}$ enhancement. It is also interesting to note that $T_{\mathrm{c}} \approx$ 3 K has been found in other studies where TaS$_2$ crystals were subject to oxidative environments such as H$_2$O\cite{doi:10.1021/jacs.7b00216} or acids \cite{Peng2018}. As shown by our calculations, oxygenation-driven enhancement of superconductivity can be explained solely by the enhanced electron-phonon coupling (see Fig.~\ref{fig:Fig4}) that may also explain why ultrathin crystals show very similar  maximum  $T_{\mathrm{c}}$, despite different preparation routes.  However, the possibility of an interplay between superconductivity and the CDW state cannot be ruled out, either, since defects -- even substitutional defects that restore the crystal lattice, as in our case -- can interfere with the long-range order of the CDW. A possible competition between superconductivity and CDW states may also explain why our experimental critical temperatures remain relatively modest (3 K) compared to the theoretically predicted values.

At this point we must emphasize that the oxygenation process itself depends on the sample thickness. Indeed, our ADF-STEM analysis has indicated faster oxygenation rates for thinner samples. As a result, the beneficial oxygen healing of the lattice can be expected to develop more gradually in the thicker samples, which can explain that freshly prepared 5-layer sample showed a higher carrier density and lower $T_{\mathrm{c}}$ compared to the trilayer and the two samples reached the same state only after aging [Fig.~\ref{fig:Transport}e,h--j]. In this scenario, the initial thickness dependence of the superconducting and CDW states in few-layer TaS$_2$ may simply result from the different rate of the oxygenation process in samples of different thickness. As such, the effect of lattice defects and of oxygenation on the interplay between superconductivity and the CDW state provides an exciting playground for further exploration of competing quantum phenomena.

We note that the highly-oxygenated limit in our study corresponds to an entire sulfur plane replaced by oxygen. Similar Janus structures, with two different chalcogen atoms on either side of a monolayer TMD (e.g., MoSSe \cite{Lu2017,Zhang2017}), have attracted a lot of interest in view of their structural symmetry breaking and anisotropy \cite{Lu2017,Zhang2017}, resulting electric dipole and enhanced Rashba spin-orbit interaction \cite{Lu2017,Dong2017}, and large piezoelectricity \cite{Dong2017}. This calls for further exploration of different TMDs with oxygen inclusion, Janus structures or otherwise, which may host further surprises, not only with regard to their elevated electron-phonon coupling, but more generally also to their optical and spintronic properties, including Ising superconductivity \cite{Lu1353,Saito2015,Xi2016,doi:10.1021/acs.nanolett.7b03026}.

With such a broad outlook to emergent quantum phenomena, our results identify controlled oxygenation as a promising route to tailor and enhance the collective quantum properties of ultrathin TMDs and their heterostructures. 

\section{Methods}

\subsection{Density functional (perturbation) theory calculations} Our density functional theory (DFT) calculations make use of the Perdew-Burke-Ernzerhof (PBE) functional implemented within the ABINIT code \cite{Gonze20092582}. To include spin-orbit coupling, fully relativistic Goedecker pseudopotentials were used \cite{PhysRevB.54.1703,Krack2005}. Here, Ta-5d$^3$6s$^2$, S-3s$^2$3p$^4$ and O-2s$^2$2p$^4$ states were included as valence electrons, together with an energy cutoff of 50 Ha for the planewave basis. To simulate the atomically thin films, 25 \AA~of vacuum was included in the unit cells. All crystal structures were relaxed so that forces on each atom were below 1 meV/\AA. A $24 \times 24 \times 1$ ($12 \times 12 \times 1$) $\Gamma$-centered Monkhorst-Pack \textbf{k}-point grid was selected for simple unit cells ($2 \times 2 \times 1$ supercells). To calculate the defect formation energies, chemical potentials of the elemental phases were calculated for body-centered cubic Ta (spacegroup Im$\bar{3}$m), rhombohedral S$_{18}$ (spacegroup R$\bar{3}$) and molecular O$_2$. Our density functional perturbation theory (DFPT) calculations employed a $24 \times 24 \times 1$ ($12 \times 12 \times 1$) $\textbf{k}$-point grid and a $12 \times 12 \times 1$ ($6 \times 6 \times 1$) $\textbf{q}$-point grid for simple unit cells (the $2 \times 2 \times 1$ supercell used for Ta$_4$S$_7$O). For an accurate description of TaSO, an increased cutoff energy of 65 Ha and a $36 \times 36 \times 1$ \textbf{k}-point grid were used. 

\subsection{Device fabrication} TaS$_2$ samples were first exfoliated onto a SiO$_2$/Si substrate coated with poly-\\propylene carbonate (PPC) and flakes of suitable thickness were identified with optical microscopy. Subsequently, encapsulated hBN/TaS$_2$/hBN (graphene/TaS$_2$/graphene) heterostructures for transport (STEM and EELS) measurements were prepared and transfered in an inert atmosphere (in an Ar filled glovebox) following the procedure described in refs.~\citenum{doi:10.1021/acs.nanolett.5b00648,Khestanova2018}. For transport measurements, selected TaS$_2$ flakes were encapsulated between two 2--5-layer thick hBN crystals (cf. Fig.~\ref{fig:Transport}a--d and Supplementary Fig.~S6). Our choice of  atomically thin top hBN allowed us to reliably determine the thickness of TaS$_2$ using atomic force microscopy mapping. After encapsulation, electrical contacts were added to the transport devices using electron beam lithography with a poly(methyl methacrylate) (PMMA) resist. For the TEM samples a layer of PMMA was spun on top of a the graphene/TaS$_2$/graphene stack to serve as support during the KOH etching of SiO$_2$ to release the sample from the substrate and transfer it onto SiN support grids. 

\subsection{ADF-STEM characterisation and EELS analysis} High-resolution STEM characterisation was carried out using a JEOL ARM300CF double aberration corrected microscope, operating at 80 kV accelerating voltage, with 67 pA beam current and a convergence semiangle of 32 mrad. Annular dark field (ADF) images were acquired at a collection angle range of 46--169 mrad. High-resolution STEM chemical characterisation was performed using an FEI Titan G2 ChemiSTEM operating at 80 kV, with a convergence semiangle of 21 mrad and fitted with a Gatan Imaging Filter (GIF) Quantum ER for electron-energy loss spectroscopy (EELS) and a Super-X, 4 SDD detector with a total collection angle of 0.7 srad for energy dispersive X-ray spectroscopy (EDXS). EELS and EDXS was performed simultaneously with a GIF collection semiangle of 62 mrad, with high loss spectra obtained for an energy range of 110--622 eV (dispersion of 0.25 eV). 

All EEL spectra and spectrum images were processed using the open source python package HyperSpy (available via {zenodo.org/record/3396791}). Individual spectra shown in Fig.~\ref{fig:EELS}a,b were acquired using an integration time of 14 s. Low-loss spectra containing the zero loss peak were acquired quasi-simultaneously with an integration time of 14 ms. All spectra were taken from the same area of the TaS$_2$ flake but we avoided probing the same spot  to avoid contamination build-up produced by the electron beam. Core loss spectra were calibrated in energy using the zero loss peak in the corresponding low loss spectra. Separate components corresponding to S L, C K, and O K core loss edges were separated from the background using least-squares fitting to model spectra. Due to the overlap of the core-loss peaks and tails from sulfur, carbon and oxygen, it was not possible to separate them by a simple power-law background fitting; the fitting parameters had to be set manually. 

Transport measurements were done with lock-in technique in four-probe configuration, and the sheet resistances were calculated according to van der Pauw's theorem\cite{vdPauw1958}.

\begin{acknowledgement}

This work was supported by Research Foundation-Flanders (FWO). J.Be. acknowledges support of a postdoctoral fellowship of the FWO. The computational resources and services used for the first-principles calculations in this work were provided by the VSC (Flemish Supercomputer Center), funded by the FWO and the Flemish Government -- department EWI. S.J.H., D.H. and S.F. would like to thank the Engineering and Physical Sciences Research Council (EPSRC) U.K (Grants EP/R031711/1, EP/P009050/1 and the Graphene NOWNANO CDT) and the European Research Council (ERC) under the European Union's Horizon 2020 research and innovation program (Grant Agreement ERC-2016-STG-EvoluTEM-715502, the Hetero2D Synergy grant) for funding.

\end{acknowledgement}

\begin{suppinfo}

Supporting information (PDF) includes (i) First-principles defect analysis, (ii) Identification of elements and defects in STEM-ADF imaging, (iii) Analysis of EELS data, and (iv) Transport measurements. 
 
\end{suppinfo}

\bibliography{biblio}

\providecommand{\latin}[1]{#1}
\makeatletter
\providecommand{\doi}
  {\begingroup\let\do\@makeother\dospecials
  \catcode`\{=1 \catcode`\}=2 \doi@aux}
\providecommand{\doi@aux}[1]{\endgroup\texttt{#1}}
\makeatother
\providecommand*\mcitethebibliography{\thebibliography}
\csname @ifundefined\endcsname{endmcitethebibliography}
  {\let\endmcitethebibliography\endthebibliography}{}
\begin{mcitethebibliography}{49}
\providecommand*\natexlab[1]{#1}
\providecommand*\mciteSetBstSublistMode[1]{}
\providecommand*\mciteSetBstMaxWidthForm[2]{}
\providecommand*\mciteBstWouldAddEndPuncttrue
  {\def\EndOfBibitem{\unskip.}}
\providecommand*\mciteBstWouldAddEndPunctfalse
  {\let\EndOfBibitem\relax}
\providecommand*\mciteSetBstMidEndSepPunct[3]{}
\providecommand*\mciteSetBstSublistLabelBeginEnd[3]{}
\providecommand*\EndOfBibitem{}
\mciteSetBstSublistMode{f}
\mciteSetBstMaxWidthForm{subitem}{(\alph{mcitesubitemcount})}
\mciteSetBstSublistLabelBeginEnd
  {\mcitemaxwidthsubitemform\space}
  {\relax}
  {\relax}

\bibitem[Anderson(1959)]{ANDERSON195926}
Anderson,~P. Theory of dirty superconductors. \emph{J. Phys. Chem. Solids}
  \textbf{1959}, \emph{11}, 26 -- 30\relax
\mciteBstWouldAddEndPuncttrue
\mciteSetBstMidEndSepPunct{\mcitedefaultmidpunct}
{\mcitedefaultendpunct}{\mcitedefaultseppunct}\relax
\EndOfBibitem
\bibitem[Uchihashi(2017)]{0953-2048-30-1-013002}
Uchihashi,~T. Two-dimensional superconductors with atomic-scale thickness.
  \emph{Supercond. Sci. Technol.} \textbf{2017}, \emph{30}, 013002\relax
\mciteBstWouldAddEndPuncttrue
\mciteSetBstMidEndSepPunct{\mcitedefaultmidpunct}
{\mcitedefaultendpunct}{\mcitedefaultseppunct}\relax
\EndOfBibitem
\bibitem[Haviland \latin{et~al.}(1989)Haviland, Liu, and
  Goldman]{PhysRevLett.62.2180}
Haviland,~D.~B.; Liu,~Y.; Goldman,~A.~M. Onset of superconductivity in the
  two-dimensional limit. \emph{Phys. Rev. Lett.} \textbf{1989}, \emph{62},
  2180--2183\relax
\mciteBstWouldAddEndPuncttrue
\mciteSetBstMidEndSepPunct{\mcitedefaultmidpunct}
{\mcitedefaultendpunct}{\mcitedefaultseppunct}\relax
\EndOfBibitem
\bibitem[Manzeli \latin{et~al.}(2017)Manzeli, Ovchinnikov, Pasquier, Yazyev,
  and Kis]{Manzeli2017}
Manzeli,~S.; Ovchinnikov,~D.; Pasquier,~D.; Yazyev,~O.~V.; Kis,~A. 2{D}
  transition metal dichalcogenides. \emph{Nat. Rev. Mater.} \textbf{2017},
  \emph{2}, 17033\relax
\mciteBstWouldAddEndPuncttrue
\mciteSetBstMidEndSepPunct{\mcitedefaultmidpunct}
{\mcitedefaultendpunct}{\mcitedefaultseppunct}\relax
\EndOfBibitem
\bibitem[Lu \latin{et~al.}(2015)Lu, Zheliuk, Leermakers, Yuan, Zeitler, Law,
  and Ye]{Lu1353}
Lu,~J.~M.; Zheliuk,~O.; Leermakers,~I.; Yuan,~N. F.~Q.; Zeitler,~U.;
  Law,~K.~T.; Ye,~J.~T. Evidence for two-dimensional {I}sing superconductivity
  in gated {M}o{S}$_2$. \emph{Science} \textbf{2015}, \emph{350},
  1353--1357\relax
\mciteBstWouldAddEndPuncttrue
\mciteSetBstMidEndSepPunct{\mcitedefaultmidpunct}
{\mcitedefaultendpunct}{\mcitedefaultseppunct}\relax
\EndOfBibitem
\bibitem[Saito \latin{et~al.}(2015)Saito, Nakamura, Bahramy, Kohama, Ye,
  Kasahara, Nakagawa, Onga, Tokunaga, Nojima, Yanase, and Iwasa]{Saito2015}
Saito,~Y.; Nakamura,~Y.; Bahramy,~M.~S.; Kohama,~Y.; Ye,~J.; Kasahara,~Y.;
  Nakagawa,~Y.; Onga,~M.; Tokunaga,~M.; Nojima,~T.; Yanase,~Y.; Iwasa,~Y.
  Superconductivity protected by spin-valley locking in ion-gated {M}o{S}$_2$.
  \emph{Nat. Phys.} \textbf{2015}, \emph{12}, 144\relax
\mciteBstWouldAddEndPuncttrue
\mciteSetBstMidEndSepPunct{\mcitedefaultmidpunct}
{\mcitedefaultendpunct}{\mcitedefaultseppunct}\relax
\EndOfBibitem
\bibitem[Xi \latin{et~al.}(2016)Xi, Wang, Zhao, Park, Law, Berger, Forro, Shan,
  and Mak]{Xi2016}
Xi,~X.; Wang,~Z.; Zhao,~W.; Park,~J.-H.; Law,~K.~T.; Berger,~H.; Forro,~L.;
  Shan,~J.; Mak,~K.~F. Ising pairing in superconducting {N}b{S}e$_2$ atomic
  layers. \emph{Nat. Phys.} \textbf{2016}, \emph{12}, 139--143\relax
\mciteBstWouldAddEndPuncttrue
\mciteSetBstMidEndSepPunct{\mcitedefaultmidpunct}
{\mcitedefaultendpunct}{\mcitedefaultseppunct}\relax
\EndOfBibitem
\bibitem[Xing \latin{et~al.}(2017)Xing, Zhao, Shan, Zheng, Zhang, Fu, Liu,
  Tian, Xi, Liu, Feng, Lin, Ji, Chen, Xue, and
  Wang]{doi:10.1021/acs.nanolett.7b03026}
Xing,~Y. \latin{et~al.}  Ising Superconductivity and Quantum Phase Transition
  in Macro-Size Monolayer {N}b{S}e$_2$. \emph{Nano Lett.} \textbf{2017},
  \emph{17}, 6802--6807\relax
\mciteBstWouldAddEndPuncttrue
\mciteSetBstMidEndSepPunct{\mcitedefaultmidpunct}
{\mcitedefaultendpunct}{\mcitedefaultseppunct}\relax
\EndOfBibitem
\bibitem[Xi \latin{et~al.}(2015)Xi, Zhao, Wang, Berger, Forr{\'o}, Shan, and
  Mak]{Xi2015}
Xi,~X.; Zhao,~L.; Wang,~Z.; Berger,~H.; Forr{\'o},~L.; Shan,~J.; Mak,~K.~F.
  Strongly enhanced charge-density-wave order in monolayer {N}b{S}e$_2$.
  \emph{Nat. Nanotechnol.} \textbf{2015}, \emph{10}, 765\relax
\mciteBstWouldAddEndPuncttrue
\mciteSetBstMidEndSepPunct{\mcitedefaultmidpunct}
{\mcitedefaultendpunct}{\mcitedefaultseppunct}\relax
\EndOfBibitem
\bibitem[Kolekar \latin{et~al.}(2018)Kolekar, Bonilla, Ma, Diaz, and
  Batzill]{2053-1583-5-1-015006}
Kolekar,~S.; Bonilla,~M.; Ma,~Y.; Diaz,~H.~C.; Batzill,~M. Layer- and
  substrate-dependent charge density wave criticality in 1{T}--{T}i{S}e$_2$.
  \emph{2D Mater.} \textbf{2018}, \emph{5}, 015006\relax
\mciteBstWouldAddEndPuncttrue
\mciteSetBstMidEndSepPunct{\mcitedefaultmidpunct}
{\mcitedefaultendpunct}{\mcitedefaultseppunct}\relax
\EndOfBibitem
\bibitem[Hovden \latin{et~al.}(2016)Hovden, Tsen, Liu, Savitzky, El~Baggari,
  Liu, Lu, Sun, Kim, Pasupathy, and Kourkoutis]{Hovden11420}
Hovden,~R.; Tsen,~A.~W.; Liu,~P.; Savitzky,~B.~H.; El~Baggari,~I.; Liu,~Y.;
  Lu,~W.; Sun,~Y.; Kim,~P.; Pasupathy,~A.~N.; Kourkoutis,~L.~F. Atomic lattice
  disorder in charge-density-wave phases of exfoliated dichalcogenides
  {1T-TaS$_2$}. \emph{Proc. Natl. Acad. Sci. U.S.A.} \textbf{2016}, \emph{113},
  11420--11424\relax
\mciteBstWouldAddEndPuncttrue
\mciteSetBstMidEndSepPunct{\mcitedefaultmidpunct}
{\mcitedefaultendpunct}{\mcitedefaultseppunct}\relax
\EndOfBibitem
\bibitem[Novello \latin{et~al.}(2017)Novello, Spera, Scarfato, Ubaldini,
  Giannini, Bowler, and Renner]{PhysRevLett.118.017002}
Novello,~A.~M.; Spera,~M.; Scarfato,~A.; Ubaldini,~A.; Giannini,~E.;
  Bowler,~D.~R.; Renner,~C. Stripe and Short Range Order in the Charge Density
  Wave of 1${T}$-{C}u$_x${T}i{S}e$_2$. \emph{Phys. Rev. Lett.} \textbf{2017},
  \emph{118}, 017002\relax
\mciteBstWouldAddEndPuncttrue
\mciteSetBstMidEndSepPunct{\mcitedefaultmidpunct}
{\mcitedefaultendpunct}{\mcitedefaultseppunct}\relax
\EndOfBibitem
\bibitem[P\'{a}sztor \latin{et~al.}(2017)P\'{a}sztor, Scarfato, Barreteau,
  Giannini, and Renner]{2053-1583-4-4-041005}
P\'{a}sztor,~A.; Scarfato,~A.; Barreteau,~C.; Giannini,~E.; Renner,~C.
  Dimensional crossover of the charge density wave transition in thin
  exfoliated {V}{S}e$_2$. \emph{2D Mater.} \textbf{2017}, \emph{4},
  041005\relax
\mciteBstWouldAddEndPuncttrue
\mciteSetBstMidEndSepPunct{\mcitedefaultmidpunct}
{\mcitedefaultendpunct}{\mcitedefaultseppunct}\relax
\EndOfBibitem
\bibitem[Yang \latin{et~al.}(2018)Yang, Fang, Fatemi, Ruhman,
  Navarro-Moratalla, Watanabe, Taniguchi, Kaxiras, and
  Jarillo-Herrero]{PhysRevB.98.035203}
Yang,~Y.; Fang,~S.; Fatemi,~V.; Ruhman,~J.; Navarro-Moratalla,~E.;
  Watanabe,~K.; Taniguchi,~T.; Kaxiras,~E.; Jarillo-Herrero,~P. Enhanced
  superconductivity upon weakening of charge density wave transport in
  $2H-\mathrm{TaS}_{2}$ in the two-dimensional limit. \emph{Phys. Rev. B}
  \textbf{2018}, \emph{98}, 035203\relax
\mciteBstWouldAddEndPuncttrue
\mciteSetBstMidEndSepPunct{\mcitedefaultmidpunct}
{\mcitedefaultendpunct}{\mcitedefaultseppunct}\relax
\EndOfBibitem
\bibitem[Navarro-Moratalla \latin{et~al.}(2016)Navarro-Moratalla, Island,
  Ma{\~n}as-Valero, Pinilla-Cienfuegos, Castellanos-Gomez, Quereda,
  Rubio-Bollinger, Chirolli, Silva-Guill{\'e}n, Agra{\"i}t, Steele, Guinea,
  van~der Zant, and Coronado]{Navarro-Moratalla2016}
Navarro-Moratalla,~E.; Island,~J.~O.; Ma{\~n}as-Valero,~S.;
  Pinilla-Cienfuegos,~E.; Castellanos-Gomez,~A.; Quereda,~J.;
  Rubio-Bollinger,~G.; Chirolli,~L.; Silva-Guill{\'e}n,~J.~A.; Agra{\"i}t,~N.;
  Steele,~G.~A.; Guinea,~F.; van~der Zant,~H. S.~J.; Coronado,~E. Enhanced
  superconductivity in atomically thin {T}a{S}$_2$. \emph{Nat. Commun.}
  \textbf{2016}, \emph{7}, 11043\relax
\mciteBstWouldAddEndPuncttrue
\mciteSetBstMidEndSepPunct{\mcitedefaultmidpunct}
{\mcitedefaultendpunct}{\mcitedefaultseppunct}\relax
\EndOfBibitem
\bibitem[de~la Barrera \latin{et~al.}(2018)de~la Barrera, Sinko, Gopalan,
  Sivadas, Seyler, Watanabe, Taniguchi, Tsen, Xu, Xiao, and
  Hunt]{delaBarrera2018}
de~la Barrera,~S.~C.; Sinko,~M.~R.; Gopalan,~D.~P.; Sivadas,~N.; Seyler,~K.~L.;
  Watanabe,~K.; Taniguchi,~T.; Tsen,~A.~W.; Xu,~X.; Xiao,~D.; Hunt,~B.~M.
  Tuning Ising superconductivity with layer and spin-orbit coupling in
  two-dimensional transition-metal dichalcogenides. \emph{Nat. Commun.}
  \textbf{2018}, \emph{9}, 1427\relax
\mciteBstWouldAddEndPuncttrue
\mciteSetBstMidEndSepPunct{\mcitedefaultmidpunct}
{\mcitedefaultendpunct}{\mcitedefaultseppunct}\relax
\EndOfBibitem
\bibitem[Pan \latin{et~al.}(2017)Pan, Guo, Song, Lai, Li, Zhao, Zhang, Mu, Bu,
  Lin, Xie, Chen, and Huang]{doi:10.1021/jacs.7b00216}
Pan,~J.; Guo,~C.; Song,~C.; Lai,~X.; Li,~H.; Zhao,~W.; Zhang,~H.; Mu,~G.;
  Bu,~K.; Lin,~T.; Xie,~X.; Chen,~M.; Huang,~F. Enhanced Superconductivity in
  Restacked {T}a{S}$_2$ Nanosheets. \emph{J. Am. Chem. Soc.} \textbf{2017},
  \emph{139}, 4623--4626\relax
\mciteBstWouldAddEndPuncttrue
\mciteSetBstMidEndSepPunct{\mcitedefaultmidpunct}
{\mcitedefaultendpunct}{\mcitedefaultseppunct}\relax
\EndOfBibitem
\bibitem[Meyer \latin{et~al.}(1975)Meyer, Howard, Stewart, Acrivos, and
  Geballe]{doi:10.1063/1.430342}
Meyer,~S.~F.; Howard,~R.~E.; Stewart,~G.~R.; Acrivos,~J.~V.; Geballe,~T.~H.
  Properties of intercalated 2H-NbSe$_2$, 4Hb-TaS$_2$, and 1T-TaS$_2$. \emph{J.
  Chem. Phys.} \textbf{1975}, \emph{62}, 4411--4419\relax
\mciteBstWouldAddEndPuncttrue
\mciteSetBstMidEndSepPunct{\mcitedefaultmidpunct}
{\mcitedefaultendpunct}{\mcitedefaultseppunct}\relax
\EndOfBibitem
\bibitem[Wang \latin{et~al.}(2018)Wang, Sun, Abdelwahab, Cao, Yu, Ju, Zhu, Fu,
  Chu, Xu, and Loh]{Wang2018}
Wang,~Z.; Sun,~Y.-Y.; Abdelwahab,~I.; Cao,~L.; Yu,~W.; Ju,~H.; Zhu,~J.; Fu,~W.;
  Chu,~L.; Xu,~H.; Loh,~K.~P. Surface-Limited Superconducting Phase Transition
  on {1T-TaS$_2$}. \emph{ACS Nano} \textbf{2018}, \emph{12}, 12619--12628\relax
\mciteBstWouldAddEndPuncttrue
\mciteSetBstMidEndSepPunct{\mcitedefaultmidpunct}
{\mcitedefaultendpunct}{\mcitedefaultseppunct}\relax
\EndOfBibitem
\bibitem[Khestanova \latin{et~al.}(2018)Khestanova, Birkbeck, Zhu, Cao, Yu,
  Ghazaryan, Yin, Berger, Forr{\'o}, Taniguchi, Watanabe, Gorbachev,
  Mishchenko, Geim, and Grigorieva]{Khestanova2018}
Khestanova,~E.; Birkbeck,~J.; Zhu,~M.; Cao,~Y.; Yu,~G.~L.; Ghazaryan,~D.;
  Yin,~J.; Berger,~H.; Forr{\'o},~L.; Taniguchi,~T.; Watanabe,~K.;
  Gorbachev,~R.~V.; Mishchenko,~A.; Geim,~A.~K.; Grigorieva,~I.~V. Unusual
  Suppression of the Superconducting Energy Gap and Critical Temperature in
  Atomically Thin {NbSe$_2$}. \emph{Nano Lett.} \textbf{2018}, \emph{18},
  2623--2629\relax
\mciteBstWouldAddEndPuncttrue
\mciteSetBstMidEndSepPunct{\mcitedefaultmidpunct}
{\mcitedefaultendpunct}{\mcitedefaultseppunct}\relax
\EndOfBibitem
\bibitem[Bekaert \latin{et~al.}(2019)Bekaert, Petrov, Aperis, Oppeneer, and
  Milo\ifmmode \check{s}\else \v{s}\fi{}evi\ifmmode~\acute{c}\else
  \'{c}\fi{}]{PhysRevLett.123.077001}
Bekaert,~J.; Petrov,~M.; Aperis,~A.; Oppeneer,~P.~M.; Milo\ifmmode
  \check{s}\else \v{s}\fi{}evi\ifmmode~\acute{c}\else \'{c}\fi{},~M.~V.
  Hydrogen-Induced High-Temperature Superconductivity in Two-Dimensional
  Materials: The Example of Hydrogenated Monolayer {MgB}$_2$. \emph{Phys. Rev.
  Lett.} \textbf{2019}, \emph{123}, 077001\relax
\mciteBstWouldAddEndPuncttrue
\mciteSetBstMidEndSepPunct{\mcitedefaultmidpunct}
{\mcitedefaultendpunct}{\mcitedefaultseppunct}\relax
\EndOfBibitem
\bibitem[Gonze \latin{et~al.}(2009)Gonze, Amadon, Anglade, Beuken, Bottin,
  Boulanger, Bruneval, Caliste, Caracas, C\^{o}t\'{e}, Deutsch, Genovese,
  Ghosez, Giantomassi, Goedecker, Hamann, Hermet, Jollet, Jomard, Leroux,
  Mancini, Mazevet, Oliveira, Onida, Pouillon, Rangel, Rignanese, Sangalli,
  Shaltaf, Torrent, Verstraete, Zerah, and Zwanziger]{Gonze20092582}
Gonze,~X. \latin{et~al.}  ABINIT: First-principles approach to material and
  nanosystem properties. \emph{Comput. Phys. Commun.} \textbf{2009},
  \emph{180}, 2582 -- 2615\relax
\mciteBstWouldAddEndPuncttrue
\mciteSetBstMidEndSepPunct{\mcitedefaultmidpunct}
{\mcitedefaultendpunct}{\mcitedefaultseppunct}\relax
\EndOfBibitem
\bibitem[Lin \latin{et~al.}(2019)Lin, Zhu, Shu, Lin, Xu, Huang, Shi, Xi, Wang,
  and Gao]{Lin2019}
Lin,~H.; Zhu,~Q.; Shu,~D.; Lin,~D.; Xu,~J.; Huang,~X.; Shi,~W.; Xi,~X.;
  Wang,~J.; Gao,~L. Growth of environmentally stable transition metal selenide
  films. \emph{Nat. Mater.} \textbf{2019}, \emph{18}, 602--607\relax
\mciteBstWouldAddEndPuncttrue
\mciteSetBstMidEndSepPunct{\mcitedefaultmidpunct}
{\mcitedefaultendpunct}{\mcitedefaultseppunct}\relax
\EndOfBibitem
\bibitem[Peng \latin{et~al.}(2018)Peng, Yu, Wu, Zhou, Guo, Li, Zhao, Wu, and
  Xie]{Peng2018}
Peng,~J.; Yu,~Z.; Wu,~J.; Zhou,~Y.; Guo,~Y.; Li,~Z.; Zhao,~J.; Wu,~C.; Xie,~Y.
  Disorder Enhanced Superconductivity toward {TaS$_2$} Monolayer. \emph{ACS
  Nano} \textbf{2018}, \emph{12}, 9461--9466\relax
\mciteBstWouldAddEndPuncttrue
\mciteSetBstMidEndSepPunct{\mcitedefaultmidpunct}
{\mcitedefaultendpunct}{\mcitedefaultseppunct}\relax
\EndOfBibitem
\bibitem[Baroni \latin{et~al.}(1987)Baroni, Giannozzi, and
  Testa]{PhysRevLett.58.1861}
Baroni,~S.; Giannozzi,~P.; Testa,~A. Green's-function approach to linear
  response in solids. \emph{Phys. Rev. Lett.} \textbf{1987}, \emph{58},
  1861--1864\relax
\mciteBstWouldAddEndPuncttrue
\mciteSetBstMidEndSepPunct{\mcitedefaultmidpunct}
{\mcitedefaultendpunct}{\mcitedefaultseppunct}\relax
\EndOfBibitem
\bibitem[Gonze \latin{et~al.}(1992)Gonze, Allan, and
  Teter]{PhysRevLett.68.3603}
Gonze,~X.; Allan,~D.~C.; Teter,~M.~P. Dielectric tensor, effective charges, and
  phonons in \ensuremath{\alpha}-quartz by variational density-functional
  perturbation theory. \emph{Phys. Rev. Lett.} \textbf{1992}, \emph{68},
  3603--3606\relax
\mciteBstWouldAddEndPuncttrue
\mciteSetBstMidEndSepPunct{\mcitedefaultmidpunct}
{\mcitedefaultendpunct}{\mcitedefaultseppunct}\relax
\EndOfBibitem
\bibitem[Savrasov(1992)]{PhysRevLett.69.2819}
Savrasov,~S.~Y. Linear response calculations of lattice dynamics using
  muffin-tin basis sets. \emph{Phys. Rev. Lett.} \textbf{1992}, \emph{69},
  2819--2822\relax
\mciteBstWouldAddEndPuncttrue
\mciteSetBstMidEndSepPunct{\mcitedefaultmidpunct}
{\mcitedefaultendpunct}{\mcitedefaultseppunct}\relax
\EndOfBibitem
\bibitem[Eliashberg(1960)]{Eliashberg1960}
Eliashberg,~G.~M. Interactions between Electrons and Lattice Vibrations in a
  Superconductor. \emph{JETP} \textbf{1960}, \emph{11}, 696\relax
\mciteBstWouldAddEndPuncttrue
\mciteSetBstMidEndSepPunct{\mcitedefaultmidpunct}
{\mcitedefaultendpunct}{\mcitedefaultseppunct}\relax
\EndOfBibitem
\bibitem[Eliashberg(1961)]{Eliashberg1961}
Eliashberg,~G.~M. Temperature Green's Function for Electrons in a
  Superconductor. \emph{JETP} \textbf{1961}, \emph{12}, 1000\relax
\mciteBstWouldAddEndPuncttrue
\mciteSetBstMidEndSepPunct{\mcitedefaultmidpunct}
{\mcitedefaultendpunct}{\mcitedefaultseppunct}\relax
\EndOfBibitem
\bibitem[Giustino(2017)]{RevModPhys.89.015003}
Giustino,~F. Electron-phonon interactions from first principles. \emph{Rev.
  Mod. Phys.} \textbf{2017}, \emph{89}, 015003\relax
\mciteBstWouldAddEndPuncttrue
\mciteSetBstMidEndSepPunct{\mcitedefaultmidpunct}
{\mcitedefaultendpunct}{\mcitedefaultseppunct}\relax
\EndOfBibitem
\bibitem[Albertini \latin{et~al.}(2017)Albertini, Liu, and
  Calandra]{PhysRevB.95.235121}
Albertini,~O.~R.; Liu,~A.~Y.; Calandra,~M. Effect of electron doping on lattice
  instabilities in single-layer 1{H}-{T}a{S}$_2$. \emph{Phys. Rev. B}
  \textbf{2017}, \emph{95}, 235121\relax
\mciteBstWouldAddEndPuncttrue
\mciteSetBstMidEndSepPunct{\mcitedefaultmidpunct}
{\mcitedefaultendpunct}{\mcitedefaultseppunct}\relax
\EndOfBibitem
\bibitem[McMillan(1968)]{PhysRev.167.331}
McMillan,~W.~L. Transition Temperature of Strong-Coupled Superconductors.
  \emph{Phys. Rev.} \textbf{1968}, \emph{167}, 331--344\relax
\mciteBstWouldAddEndPuncttrue
\mciteSetBstMidEndSepPunct{\mcitedefaultmidpunct}
{\mcitedefaultendpunct}{\mcitedefaultseppunct}\relax
\EndOfBibitem
\bibitem[Allen and Dynes(1975)Allen, and Dynes]{PhysRevB.12.905}
Allen,~P.~B.; Dynes,~R.~C. Transition temperature of strong-coupled
  superconductors reanalyzed. \emph{Phys. Rev. B} \textbf{1975}, \emph{12},
  905--922\relax
\mciteBstWouldAddEndPuncttrue
\mciteSetBstMidEndSepPunct{\mcitedefaultmidpunct}
{\mcitedefaultendpunct}{\mcitedefaultseppunct}\relax
\EndOfBibitem
\bibitem[Morel and Anderson(1962)Morel, and Anderson]{PhysRev.125.1263}
Morel,~P.; Anderson,~P.~W. Calculation of the Superconducting State Parameters
  with Retarded Electron-Phonon Interaction. \emph{Phys. Rev.} \textbf{1962},
  \emph{125}, 1263--1271\relax
\mciteBstWouldAddEndPuncttrue
\mciteSetBstMidEndSepPunct{\mcitedefaultmidpunct}
{\mcitedefaultendpunct}{\mcitedefaultseppunct}\relax
\EndOfBibitem
\bibitem[Allen and Mitrovi\'{c}(1983)Allen, and Mitrovi\'{c}]{ALLEN19831}
Allen,~P.~B.; Mitrovi\'{c},~B. In \emph{Theory of Superconducting
  ${T}_{\mathrm{c}}$}; Ehrenreich,~H., Seitz,~F., Turnbull,~D., Eds.; Solid
  State Physics; Academic Press, 1983; Vol.~37; pp 1 -- 92\relax
\mciteBstWouldAddEndPuncttrue
\mciteSetBstMidEndSepPunct{\mcitedefaultmidpunct}
{\mcitedefaultendpunct}{\mcitedefaultseppunct}\relax
\EndOfBibitem
\bibitem[Clark \latin{et~al.}(2018)Clark, Nguyen, Hamer, Schedin, Lewis,
  Prestat, Garner, Cao, Zhu, Kashtiban, Sloan, Kepaptsoglou, Gorbachev, and
  Haigh]{Clark2018}
Clark,~N.; Nguyen,~L.; Hamer,~M.~J.; Schedin,~F.; Lewis,~E.~A.; Prestat,~E.;
  Garner,~A.; Cao,~Y.; Zhu,~M.; Kashtiban,~R.; Sloan,~J.; Kepaptsoglou,~D.;
  Gorbachev,~R.~V.; Haigh,~S.~J. Scalable Patterning of Encapsulated Black
  Phosphorus. \emph{Nano Lett.} \textbf{2018}, \emph{18}, 5373--5381\relax
\mciteBstWouldAddEndPuncttrue
\mciteSetBstMidEndSepPunct{\mcitedefaultmidpunct}
{\mcitedefaultendpunct}{\mcitedefaultseppunct}\relax
\EndOfBibitem
\bibitem[Cao \latin{et~al.}(2015)Cao, Mishchenko, Yu, Khestanova, Rooney,
  Prestat, Kretinin, Blake, Shalom, Woods, Chapman, Balakrishnan, Grigorieva,
  Novoselov, Piot, Potemski, Watanabe, Taniguchi, Haigh, Geim, and
  Gorbachev]{doi:10.1021/acs.nanolett.5b00648}
Cao,~Y. \latin{et~al.}  Quality Heterostructures from Two-Dimensional Crystals
  Unstable in Air by Their Assembly in Inert Atmosphere. \emph{Nano Lett.}
  \textbf{2015}, \emph{15}, 4914--4921\relax
\mciteBstWouldAddEndPuncttrue
\mciteSetBstMidEndSepPunct{\mcitedefaultmidpunct}
{\mcitedefaultendpunct}{\mcitedefaultseppunct}\relax
\EndOfBibitem
\bibitem[Kretinin \latin{et~al.}(2014)Kretinin, Cao, Tu, Yu, Jalil, Novoselov,
  Haigh, Gholinia, Mishchenko, Lozada, Georgiou, Woods, Withers, Blake, Eda,
  Wirsig, Hucho, Watanabe, Taniguchi, Geim, and Gorbachev]{Kretinin2014}
Kretinin,~A.~V. \latin{et~al.}  Electronic Properties of Graphene Encapsulated
  with Different Two-Dimensional Atomic Crystals. \emph{Nano Lett.}
  \textbf{2014}, \emph{14}, 3270--3276\relax
\mciteBstWouldAddEndPuncttrue
\mciteSetBstMidEndSepPunct{\mcitedefaultmidpunct}
{\mcitedefaultendpunct}{\mcitedefaultseppunct}\relax
\EndOfBibitem
\bibitem[Komsa \latin{et~al.}(2013)Komsa, Kurasch, Lehtinen, Kaiser, and
  Krasheninnikov]{PhysRevB.88.035301}
Komsa,~H.-P.; Kurasch,~S.; Lehtinen,~O.; Kaiser,~U.; Krasheninnikov,~A.~V. From
  point to extended defects in two-dimensional {M}o{S}${}_{2}$: Evolution of
  atomic structure under electron irradiation. \emph{Phys. Rev. B}
  \textbf{2013}, \emph{88}, 035301\relax
\mciteBstWouldAddEndPuncttrue
\mciteSetBstMidEndSepPunct{\mcitedefaultmidpunct}
{\mcitedefaultendpunct}{\mcitedefaultseppunct}\relax
\EndOfBibitem
\bibitem[Wang \latin{et~al.}(2016)Wang, Lee, Lee, Yoon, and Warner]{Wang2016}
Wang,~S.; Lee,~G.-D.; Lee,~S.; Yoon,~E.; Warner,~J.~H. Detailed Atomic
  Reconstruction of Extended Line Defects in Monolayer {M}o{S}${}_{2}$.
  \emph{ACS Nano} \textbf{2016}, \emph{10}, 5419--5430\relax
\mciteBstWouldAddEndPuncttrue
\mciteSetBstMidEndSepPunct{\mcitedefaultmidpunct}
{\mcitedefaultendpunct}{\mcitedefaultseppunct}\relax
\EndOfBibitem
\bibitem[Schaltin \latin{et~al.}(2012)Schaltin, D{'}Urzo, Zhao, Vantomme,
  Plank, Kothleitner, Gspan, Binnemans, and Fransaer]{C2CP41786C}
Schaltin,~S.; D{'}Urzo,~L.; Zhao,~Q.; Vantomme,~A.; Plank,~H.; Kothleitner,~G.;
  Gspan,~C.; Binnemans,~K.; Fransaer,~J. Direct electroplating of copper on
  tantalum from ionic liquids in high vacuum: origin of the tantalum oxide
  layer. \emph{Phys. Chem. Chem. Phys.} \textbf{2012}, \emph{14},
  13624--13629\relax
\mciteBstWouldAddEndPuncttrue
\mciteSetBstMidEndSepPunct{\mcitedefaultmidpunct}
{\mcitedefaultendpunct}{\mcitedefaultseppunct}\relax
\EndOfBibitem
\bibitem[Song \latin{et~al.}(2017)Song, Park, Yoon, Yoon, Kwon, Noh,
  Lansalot-Matras, Gatineau, Lee, Gautam, Cho, Lee, and Hwang]{Song2017}
Song,~S.~J.; Park,~T.; Yoon,~K.~J.; Yoon,~J.~H.; Kwon,~D.~E.; Noh,~W.;
  Lansalot-Matras,~C.; Gatineau,~S.; Lee,~H.-K.; Gautam,~S.; Cho,~D.-Y.;
  Lee,~S.~W.; Hwang,~C.~S. Comparison of the Atomic Layer Deposition of
  Tantalum Oxide Thin Films Using {T}a({N}$^t${B}u)({N}Et$_2$)$_3$,
  {T}a({N}$^t${B}u)({N}Et$_2$)$_2$Cp, and {H}$_2${O}. \emph{ACS Appl. Mater.
  Interfaces} \textbf{2017}, \emph{9}, 537--547\relax
\mciteBstWouldAddEndPuncttrue
\mciteSetBstMidEndSepPunct{\mcitedefaultmidpunct}
{\mcitedefaultendpunct}{\mcitedefaultseppunct}\relax
\EndOfBibitem
\bibitem[van~der Pauw(1958)]{vdPauw1958}
van~der Pauw,~L.~J. A Method of Measuring the Resistivity and {H}all
  Coefficient on Lamellae of Arbitrary Shape. \emph{Philips Tech. Rev.}
  \textbf{1958}, \emph{20}, 220\relax
\mciteBstWouldAddEndPuncttrue
\mciteSetBstMidEndSepPunct{\mcitedefaultmidpunct}
{\mcitedefaultendpunct}{\mcitedefaultseppunct}\relax
\EndOfBibitem
\bibitem[Lu \latin{et~al.}(2017)Lu, Zhu, Xiao, Chuu, Han, Chiu, Cheng, Yang,
  Wei, Yang, Wang, Sokaras, Nordlund, Yang, Muller, Chou, Zhang, and
  Li]{Lu2017}
Lu,~A.-Y. \latin{et~al.}  Janus monolayers of transition metal dichalcogenides.
  \emph{Nat. Nanotechnol.} \textbf{2017}, \emph{12}, 744\relax
\mciteBstWouldAddEndPuncttrue
\mciteSetBstMidEndSepPunct{\mcitedefaultmidpunct}
{\mcitedefaultendpunct}{\mcitedefaultseppunct}\relax
\EndOfBibitem
\bibitem[Zhang \latin{et~al.}(2017)Zhang, Jia, Kholmanov, Dong, Er, Chen, Guo,
  Jin, Shenoy, Shi, and Lou]{Zhang2017}
Zhang,~J.; Jia,~S.; Kholmanov,~I.; Dong,~L.; Er,~D.; Chen,~W.; Guo,~H.;
  Jin,~Z.; Shenoy,~V.~B.; Shi,~L.; Lou,~J. Janus Monolayer Transition-Metal
  Dichalcogenides. \emph{ACS Nano} \textbf{2017}, \emph{11}, 8192--8198\relax
\mciteBstWouldAddEndPuncttrue
\mciteSetBstMidEndSepPunct{\mcitedefaultmidpunct}
{\mcitedefaultendpunct}{\mcitedefaultseppunct}\relax
\EndOfBibitem
\bibitem[Dong \latin{et~al.}(2017)Dong, Lou, and Shenoy]{Dong2017}
Dong,~L.; Lou,~J.; Shenoy,~V.~B. Large In-Plane and Vertical Piezoelectricity
  in {J}anus Transition Metal Dichalchogenides. \emph{ACS Nano} \textbf{2017},
  \emph{11}, 8242--8248\relax
\mciteBstWouldAddEndPuncttrue
\mciteSetBstMidEndSepPunct{\mcitedefaultmidpunct}
{\mcitedefaultendpunct}{\mcitedefaultseppunct}\relax
\EndOfBibitem
\bibitem[Goedecker \latin{et~al.}(1996)Goedecker, Teter, and
  Hutter]{PhysRevB.54.1703}
Goedecker,~S.; Teter,~M.; Hutter,~J. Separable dual-space {G}aussian
  pseudopotentials. \emph{Phys. Rev. B} \textbf{1996}, \emph{54},
  1703--1710\relax
\mciteBstWouldAddEndPuncttrue
\mciteSetBstMidEndSepPunct{\mcitedefaultmidpunct}
{\mcitedefaultendpunct}{\mcitedefaultseppunct}\relax
\EndOfBibitem
\bibitem[Krack(2005)]{Krack2005}
Krack,~M. Pseudopotentials for {H} to {K}r optimized for gradient-corrected
  exchange-correlation functionals. \emph{Theor. Chem. Acc.} \textbf{2005},
  \emph{114}, 145--152\relax
\mciteBstWouldAddEndPuncttrue
\mciteSetBstMidEndSepPunct{\mcitedefaultmidpunct}
{\mcitedefaultendpunct}{\mcitedefaultseppunct}\relax
\EndOfBibitem
\end{mcitethebibliography}

\end{document}